\documentclass[]{elsarticle}

\usepackage{amsmath,mathtools}
\usepackage{tabularx,multirow}
\usepackage{graphicx,subfigure,comment}
\usepackage[table]{xcolor}
\usepackage{url} %for \path
\usepackage{psfrag}
\usepackage{cite}
\usepackage{dblfloatfix}

\usepackage[algoruled,medskip,dontprintsemicolon,linesnumbered,Algorithm]{algorithm}
\usepackage[noend]{algpseudocode}

\newcommand{\Fig}[1]{Fig.~\ref{fig:#1}}
\newcommand{\Sec}[1]{Sec.~\ref{sec:#1}}

\newcommand{\Alg}[1]{Alg.~\ref{alg:#1}}
\newcommand{\Line}[1]{Line~\ref{line:#1}}

\newcommand{\Bc}{\mathcal{B}}
\newcommand{\Cc}{\mathcal{C}}

\begin{document}

\begin{frontmatter}

\title{
Virtualization-based Evaluation\\
of Backhaul Performance in Vehicular Applications
%A Virtualized Testbed for Vehicular Applications
} %title

\author{Francesco Malandrino, Carla-Fabiana Chiasserini, Claudio Casetti\\
DET, Politecnico di Torino, Torino, Italy}

\begin{abstract}
Next-generation networks, based on SDN and NFV, are expected to support
a wide array of services, including vehicular safety applications. These
services come with strict delay constraints, and our goal in this paper
is to ascertain to which extent SDN/NFV-based networks are able to meet
them. To this end, we build and emulate a vehicular collision detection
system, using the popular Mininet and Docker tools, on a
real-world topology with mobility information. Using different core
network topologies and open-source SDN controllers, we measure (i) the
delay with which vehicle beacons are processed and (ii) the associated
overhead and energy consumption. We find that we can indeed meet
the latency constraints associated with vehicular safety applications,
and that SDN controllers represent a moderate contribution to the
overall energy consumption but a significant source of additional delay.
\end{abstract}

\end{frontmatter}

\section{Introduction}
\label{sec:intro}

\begin{figure}[h]
\centering
\includegraphics[width=.6\textwidth]{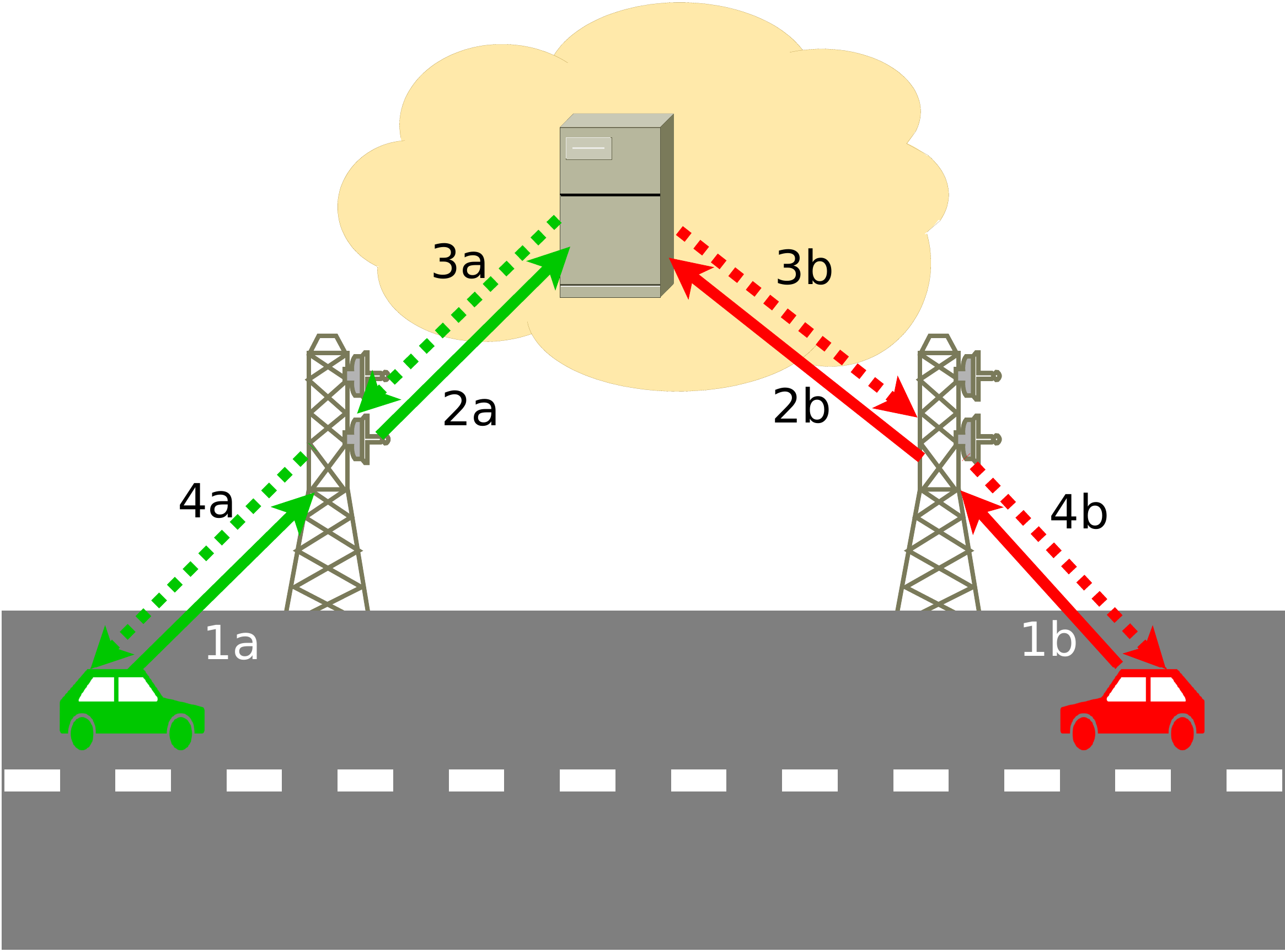}
\caption{A simple vehicular network composed of two vehicles (red and green), two road-side units (RSUs) and a centralized collision detector. Solid lines represent beacon transmissions, dashed lines correspond to collision warnings. The vehicles periodically transmit beacons (1a, 1b), which, through the RSUs, reach the collision detector (2a, 2b). The detector realizes that the vehicles are set on a collision course, and issues two collision warnings (3a, 3b) that, again through the RSUs (4a, 4b), reach the vehicles.
\label{fig:idea}
} %caption
\end{figure}

Vehicular networks are mobile wireless networks whose nodes are
represented by connected vehicles and the infrastructure supporting
them, e.g., road-side units (RSUs) providing Internet connectivity, as
exemplified in \Fig{idea}. Current and expected applications abound, and
include navigation, e.g., downloading maps or traffic updates, and
entertainment, e.g., streaming movies to on-board entertainment systems
similar to those found on airplanes.

A third, and arguably more critical, application of vehicular networks
is represented by {\em safety}: indeed, in 2015 road accidents accounted
for over 35,000~deaths in the United States alone~\citep{roads-us}, and
over one million worldwide~\citep{who-roads}. The most significant of
these safety applications is {\em collision detection}. The idea of
collision detection is fairly simple, and is summarized in \Fig{idea}.
Vehicles periodically~\citep{beacons-falko} (and anonymously~\citep{avip})
report their position, direction and speed to a {\em detector}.
The communication between vehicles and detectors happens through {\em
road-side units} (RSUs), that make communication possible even in
non-line-of-sight (NLoS) conditions, e.g., due to buildings or other
obstacles.
The detector combines these reports, determines whether any two vehicles
are set on a collision course, and, if so, it alerts their drivers.
Collision detection is especially important in presence of obstacles,
e.g., buildings, that prevent drivers from timely realizing the danger.
The importance and relevance of collision detection has been
acknowledged by transportation regulators: as recently as December 2016,
the U.S. Department of Transportation (DOT) published a Notice of
Proposed Rulemaking (NPRM) for vehicular communications~\citep{nhtsa}.
The document proposes to establish a new Federal Motor Vehicle Safety
Standard (FMVSS), No. 150, to make vehicular networking technology
compulsory: 50\%~of newly-made vehicles will have to be equipped
with such a technology in 2018, 75\% in~2019, and~100\% in~2020.

It is fairly obvious that timeliness is critical to collision detection
systems. However, satisfying latency requirements in emerging mobile network
systems, which rely on  software-defined networking (SDN) and network
function virtualization (NFV) in the backhaul (and sometimes even in the
fronthaul), may be  challenging. Indeed, while SDN
and NFV bring major improvements in terms of network flexibility and
efficiency, both imply a certain amount of overhead: such overhead is
negligible in most applications, but not when it comes to vehicular
safety.
An additional concern is represented by energy consumption: some network
nodes, e.g., solar-powered RSUs, might not be connected to a reliable
power supply; it is therefore important to know the power consumption
associated with virtual network functions (VNFs), so as to better decide
at which physical nodes to place them.

In this paper, we build, optimize, and evaluate a collision detection
system, based on Mininet and Docker, the standard tools 
for SDN emulation and containerization, respectively.
Our purpose is twofold: on the one hand, we study
the impact of SDN and NFV on the performance of vehicular networks; 
on the other, we seek to learn valuable, real-world lessons concerning
the pitfalls and implementation issues associated with our tools.

As far as the tools we use are concerned, Mininet~\citep{mininet}
recently emerged as the {\em de facto} standard for reproducible network
experiments. It emulates a full network, including software, SDN-capable
switches and virtual hosts, running arbitrary programs in separate
execution environments while sharing the file system and process space.
It is typically used in SDN research, with custom-written controllers
controlling the Mininet-emulated switches.
In our case, however, we do not write our own custom controller; rather,
we test two popular, general-purpose SDN controllers -- namely,
Pox~\citep{pox} and Floodlight~\citep{floodlight} -- and ascertain how
they impact the performance and energy consumption of our emulated
network.

In our experiments, we couple Mininet with Docker~\citep{docker}, again
the {\em de facto} standard containerization platform. Containers, often
described as lightweight virtual machines, are a virtualization
technique where applications run in isolated environments but share the
same Linux kernel, thus substantially reducing the overhead. For this
reason, they are generally viewed as the ideal way to implement network
function virtualization in next-generation networks.

The remainder of this paper is organized as follows. We start by
discussing how collision detection is carried out, in \Sec{collisions}.
Then, \Sec{testbed}  describes our reference scenario, the virtualized
network architecture, and investigates the delay over the wireless
network segment.
\Sec{placement} shows how we refine collision
detector placement, while \Sec{results} reports our findings.  Finally, we discuss related work
in \Sec{relwork} and conclude the paper in \Sec{conclusion}.

\section{Detecting collisions}
\label{sec:collisions}

Our collision detection system, depicted in \Fig{idea}, has two main
components: vehicles, and one or more {\em collision detectors}.

As specified by current standards, vehicles are in charge of
periodically sending {\em beacons}, reporting their position, direction,
and speed. In order to safeguard privacy, beacons are {\em
anonymized}~\citep{anonymity}, e.g., they do not include the vehicle
identity and report a temporary source MAC address (also called a {\em
pseudonym}~\citep{pseudo}).

The beacons are conveyed, through a set of {\em road-side units} (RSUs)
to a {\em collision detector}, running on a centralized -- and,
typically, virtualized -- server as shown in \Fig{idea}. The detector
keeps a set~$\Bc$ of {\em recently}\footnote{The beacon timeout depends
on the actual scenario; in our case we set it to one second.}
received beacons and, upon receiving a new beacon, checks it for
collisions as summarized in \Alg{collisions}. 

\begin{algorithm}[t!]
\caption{Collision detection\label{alg:collisions}
} %caption
\begin{algorithmic}[1]
\Require{$\vec{x_0},\vec{v},\Bc$} \label{line:input}
\State{$\Cc\gets\emptyset$} \label{line:init-c}
\State{$\vec{x}(t)\gets \vec{x_0}+\vec{v}t$} \label{line:pos}
\ForAll{$b\in\Bc$} \label{line:forb}
 \State{$\vec{x^b}(t)\gets \vec{x_0^b}+\vec{v^b}\cdot t$} \label{line:posb}
 \State{$\vec{d}(t)\gets \vec{x}(t)-\vec{x^b}(t)$} \label{line:dist}
 \State{$D(t)\coloneqq|\vec{d}(t)|^2\gets (\vec{v}-\vec{v^b})\cdot(\vec{v}-\vec{v^b})t^2+2(\vec{x_0}-\vec{x_0^b})\cdot(\vec{v}-\vec{v^b})t+(\vec{x_0}-\vec{x_0^b})\cdot(\vec{x_0}-\vec{x_0^b})$} \label{line:d2}
 \State{$t^{\star}\coloneqq t\colon\frac{\mathrm{d}}{\mathrm{d}t}D(t)=0\gets \frac{-(\vec{x_0}-\vec{x_0^b})\cdot(\vec{v}-\vec{v^b})}{|\vec{v}-\vec{v^b}|^2}$} \label{line:tstar}
 \If{$t^{\star}<0$} \label{line:check-past}
  \State{{\bf continue}}
 \EndIf
 \State{$d^{\star}\gets \sqrt{D(t^{\star})}$} \label{line:dstar}
 \If{$d^{\star}\leq d_{\min}$} \label{line:check-dstar}
  \State{$\Cc\gets \Cc\cup \{b\}$} \label{line:add}
 \EndIf
\EndFor
\Return{$\Cc$} \label{line:return}
\end{algorithmic}
\end{algorithm}

The algorithm, which is  based on~\citep{collisiondetection}, takes as an
input the position and speed of the current vehicle (\Line{input}),
respectively identified by vectors~$\vec{x_0}$ and~$\vec{v}$
\footnote{Note that the speed vector also includes information on the
direction.}, as well as the previous beacons in~$\Bc$. We start by
initializing the set~$\Cc$ of vehicles, with which the current vehicle
will collide, to the empty set (\Line{init-c}), and we compute how the
position of the current vehicle will change over time (\Line{pos}).
Then, for every vehicle that generated a beacon $b\in\Bc$ recently
received by the detector, we compute its position over time
(\Line{posb}) and the difference~$\vec{d}(t)$ between the positions of
such vehicle and the current vehicle (\Line{dist}). The
scalar~$D(t)\coloneqq|\vec{d}(t)|^2$, computed in \Line{d2}, represents
the square \footnote{Using the squared distance instead of the distance
itself simplifies computations.} of the distance over time. We are
interested in the minimum value that this quantity will take over time;
to this end, in \Line{tstar} we compute the time~$t^{\star}$ at
which~$D(t)$ will take its minimum value. If~$t^{\star}<0$, then the
vehicles are actually getting farther apart and no action is required
(\Line{check-past}). Otherwise, in \Line{dstar} we compute the minimum
distance~$d^{\star}$ the two vehicles will be at; if such a value is
lower than a threshold value~$d_{\min}$ (\Line{check-dstar}), then we
need to send an alert, and thus add~$b$ to~$\Cc$ (note that $b$
essentially identifies the vehicle who sent the beacon).

In summary, \Alg{collisions} returns the set~$\Cc$ of vehicles with
which the current vehicle is set to collide. This set (along with
additional information such as the time of collision) is transmitted
back to the vehicles whose  beacon was included in $\Cc$, as shown in
\Fig{idea}. The vehicles will therefore alert their drivers or, if
appropriate, directly take action, e.g., brake before the collision
happens.

\section{Network scenario and virtualized backhaul}
\label{sec:testbed}

\begin{figure}[h]
\centering
\includegraphics[width=.7\textwidth]{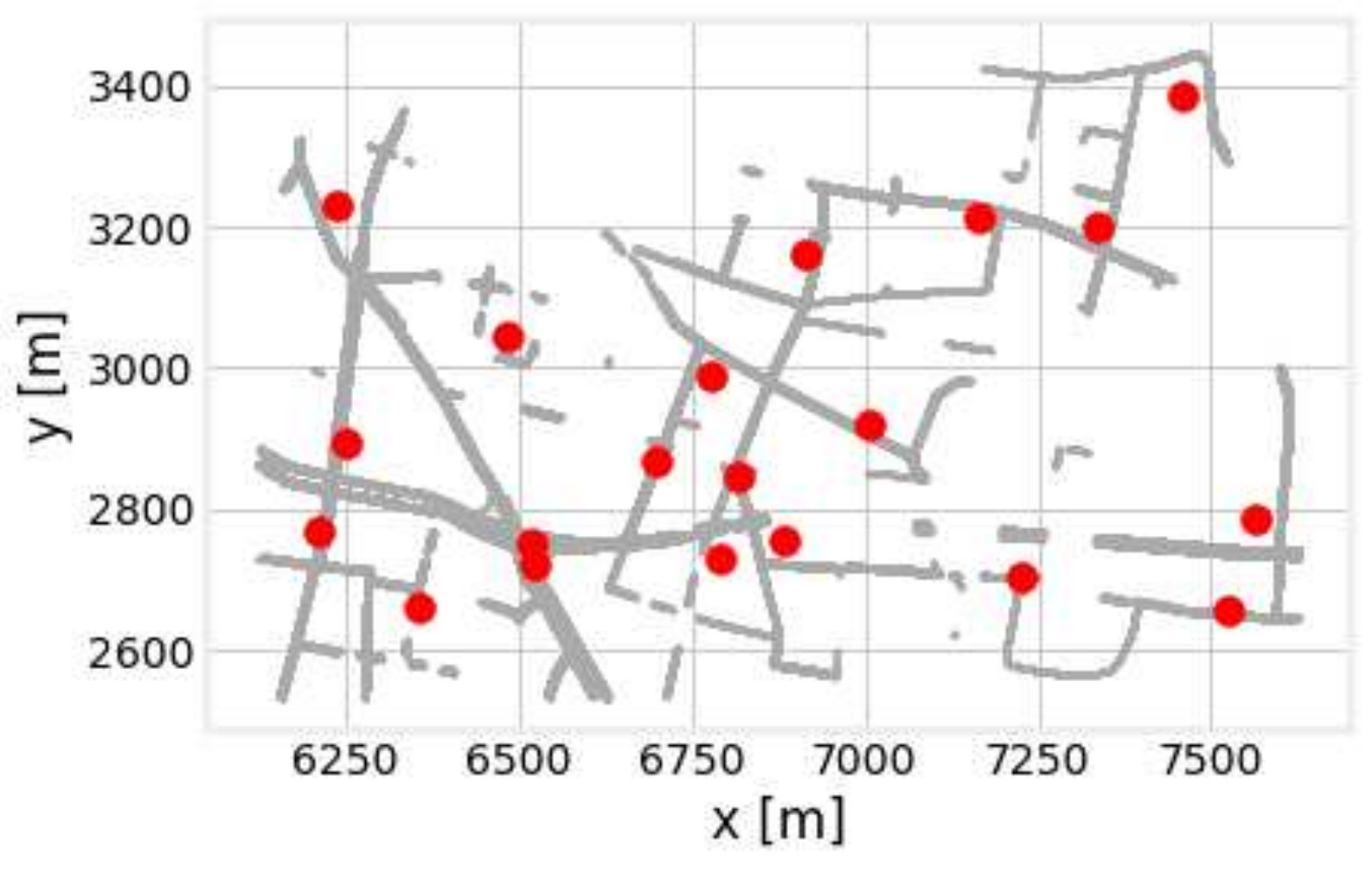}
\caption{
Road topology, with red dots corresponding to RSUs.
\label{fig:roads}
}%caption
\end{figure}

\begin{figure}[h]
\centering
\subfigure[\label{fig:trace-speed}]{
    \includegraphics[width=.48\textwidth]{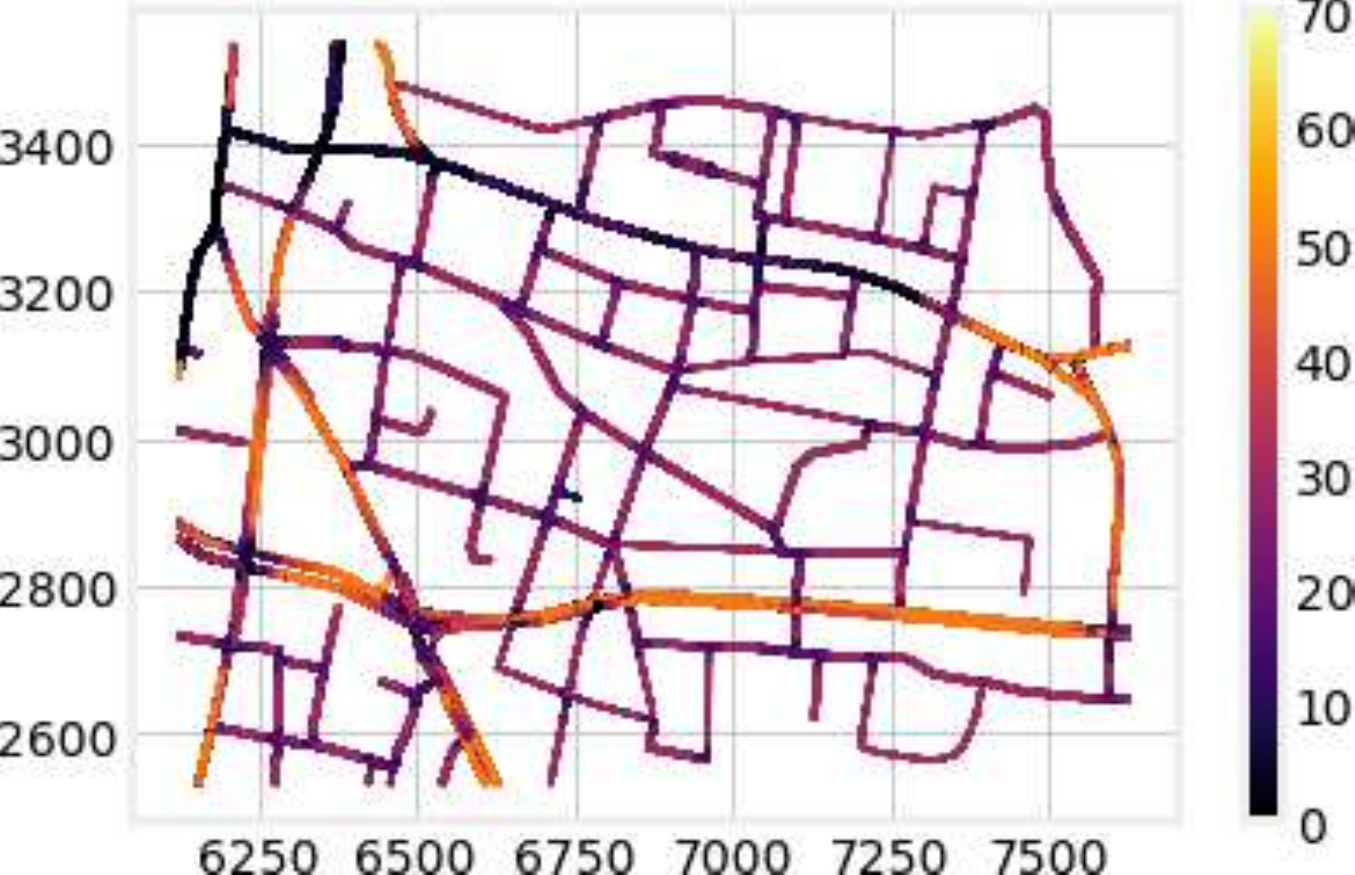}
} %subfigure
\hspace{-5mm}
\subfigure[\label{fig:trace-density}]{
    \includegraphics[width=.48\textwidth]{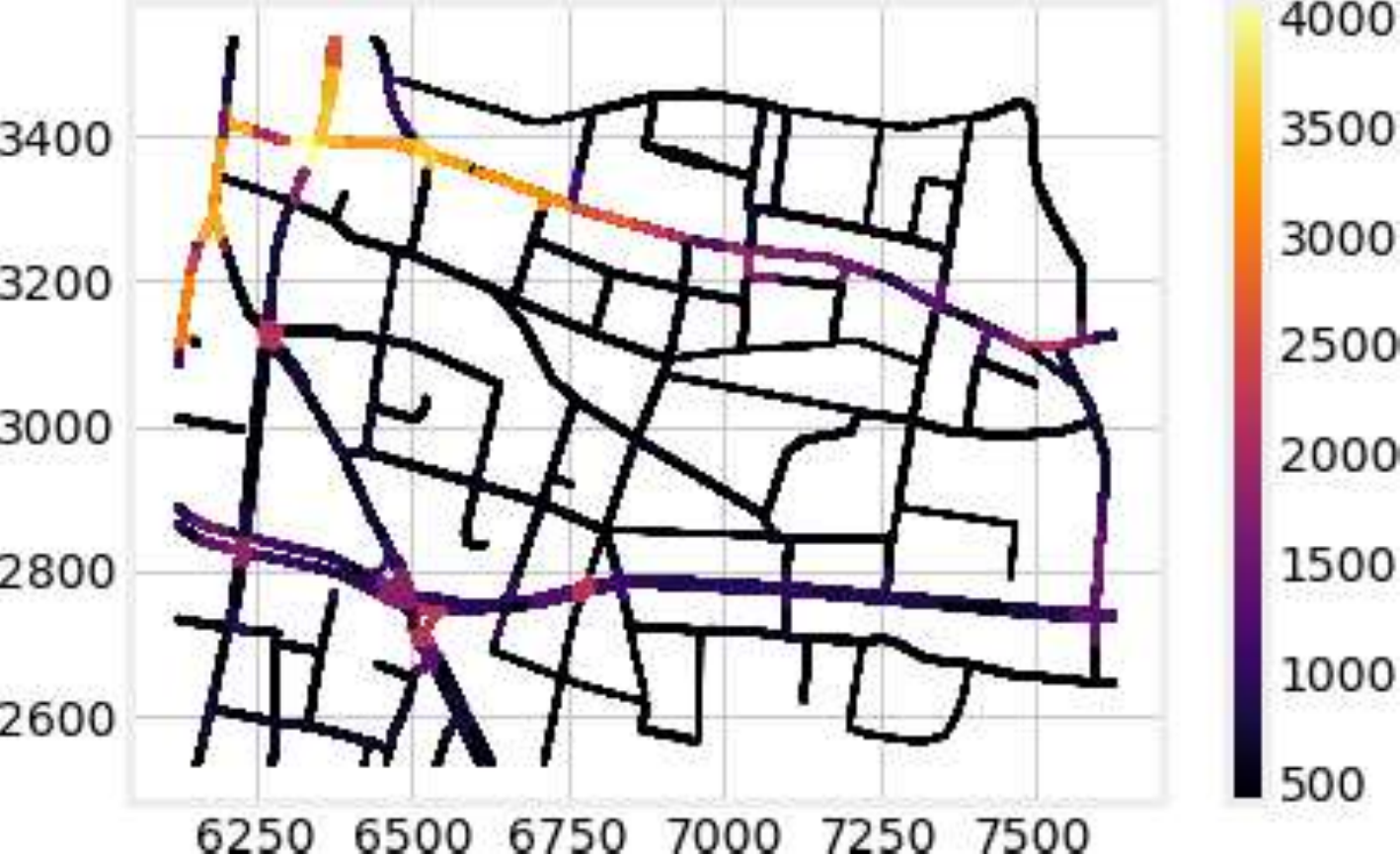}
} %subfigure
\caption{
    Speed (a) and density (b) of vehicles at different location of the trace we use. The scale is in km/h in (a) and vehicles per square kilometers in (b); darker colors correspond to higher values.
} %caption
\end{figure}

This section describes our reference network scenario and the architecture of
the virtualized bachkaul under study. Specifically,
\Sec{mobility} details the real-world, large-scale scenario we seek to
emulate, and its traffic and demand patterns. Then, in
\Sec{testbed-structure}, we discuss how we emulate such a scenario using
Mininet and Docker, as well as the applications we run within each
emulated node. Finally, \Sec{wireless} describes how the
communication on the radio link is simulated and how the 
resulting  delay is accounted for in the network emulations.

\subsection{Reference scenario}
\label{sec:mobility}

As mentioned earlier, the beacons include the position, speed and
direction of the vehicle sending them. In our experiments, this
information is obtained from the mobility trace presented
in~\citep{trace}. Therein, the authors combine a
$1.5\times 1$\,km$^2$ section of the 
real-world road
topology of the city of Ingolstadt (Germany), depicted in \Fig{roads},
and realistic vehicular mobility obtained with the SUMO
simulator~\citep{sumo}.
Ingolstadt is a medium-sized city in the Munich
  metropolitan area; the inner city includes a mixture of narrow
  streets and wider, multi-lane roads, as it is common in urban areas
  throughout the world. Taking it as our main reference scenario
  allows us to easily generalize 
  the main indications in which we are interested, to other cities and
  countries. 
It is also worth stressing that, in \Sec{results-koeln}, 
we check to which extent considering another road topology and user
mobility impacts our results.

\begin{figure}[h]
\centering
\subfigure[\label{fig:demand}]{
    \includegraphics[width=.46\textwidth]{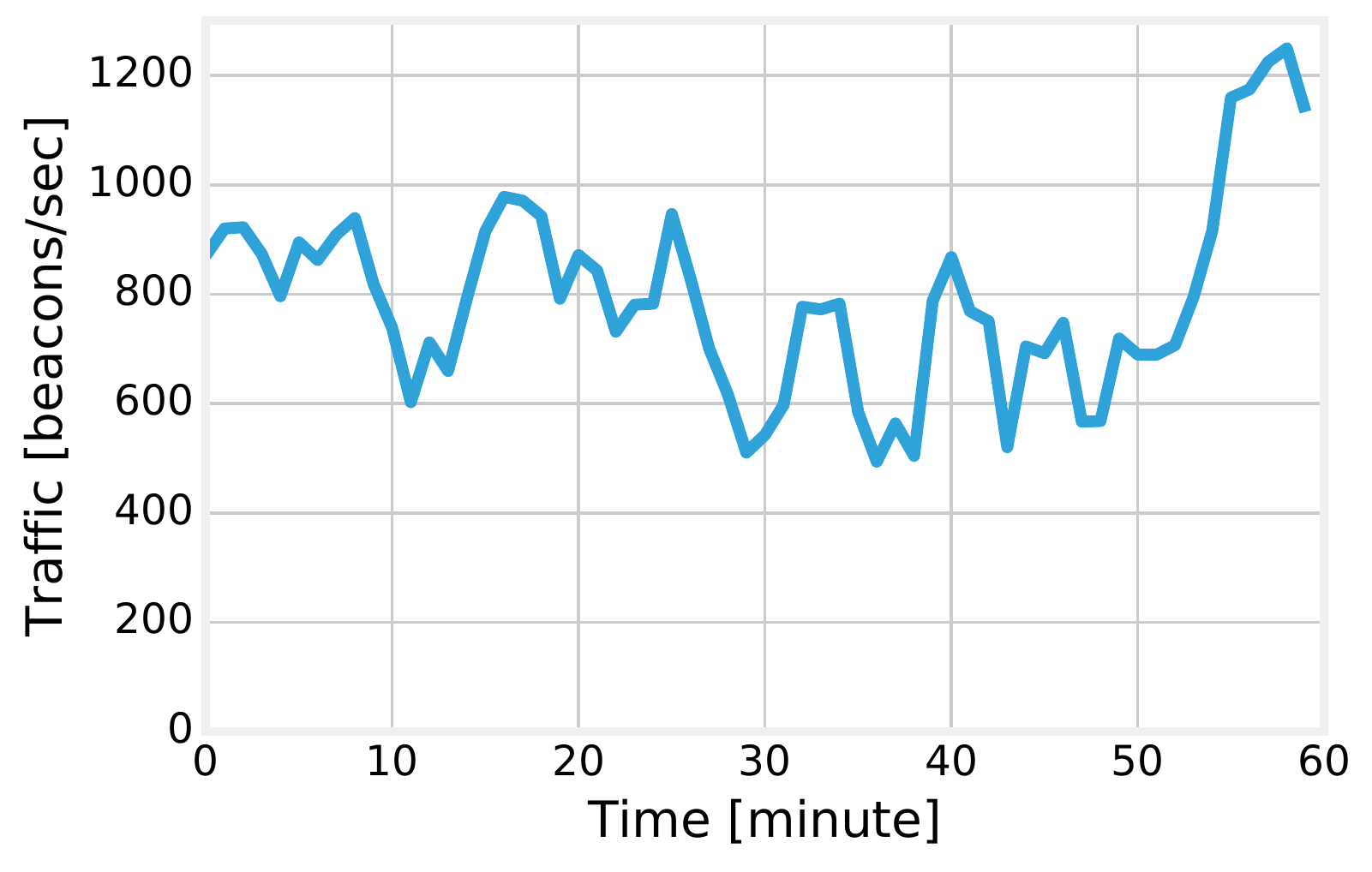}
}%subfigure
\subfigure[\label{fig:network-topo-mesh}]{
    \includegraphics[width=.46\textwidth]{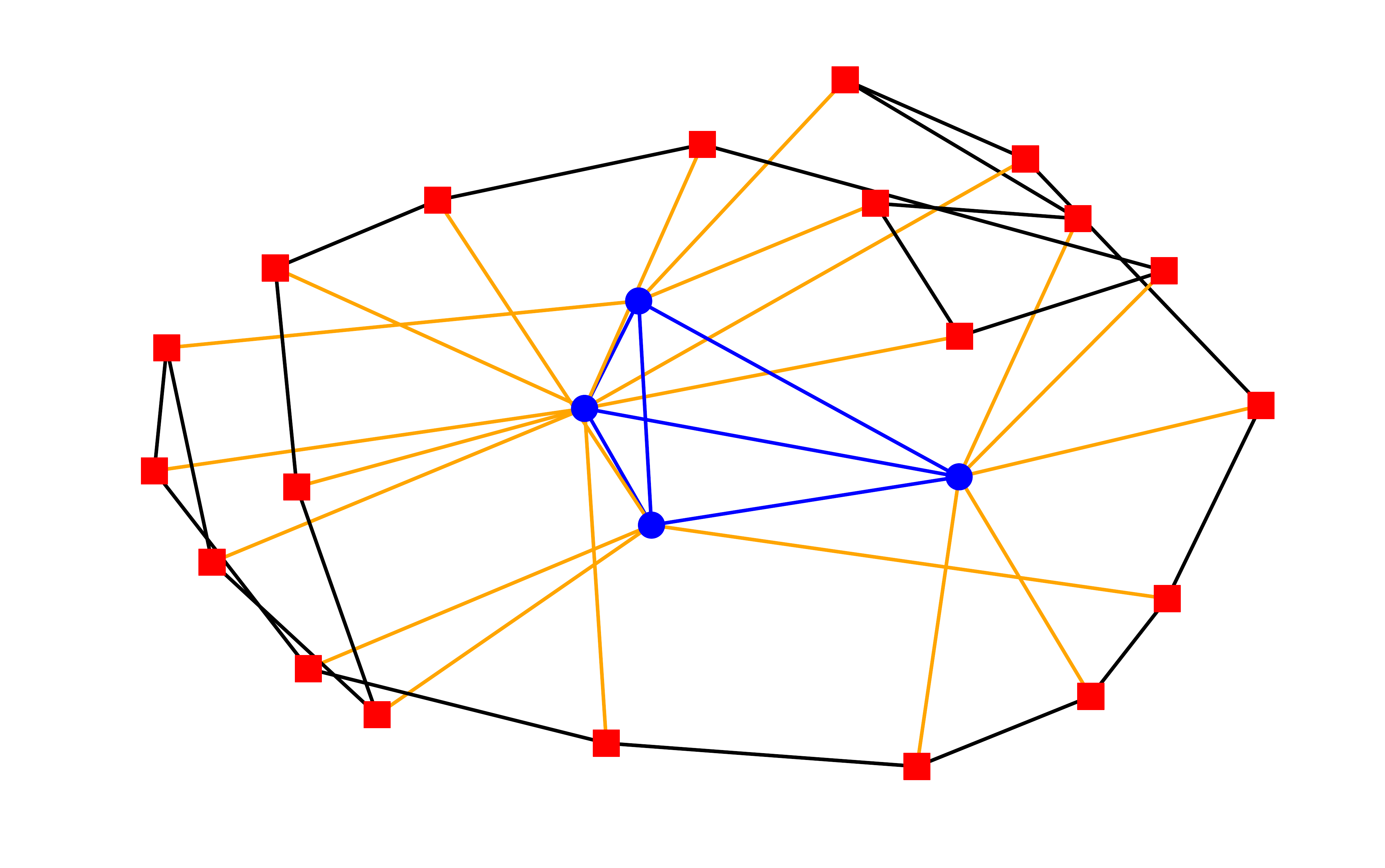}
}%subfigure
\\
\subfigure[\label{fig:network-topo-star}]{
    \includegraphics[width=.46\textwidth]{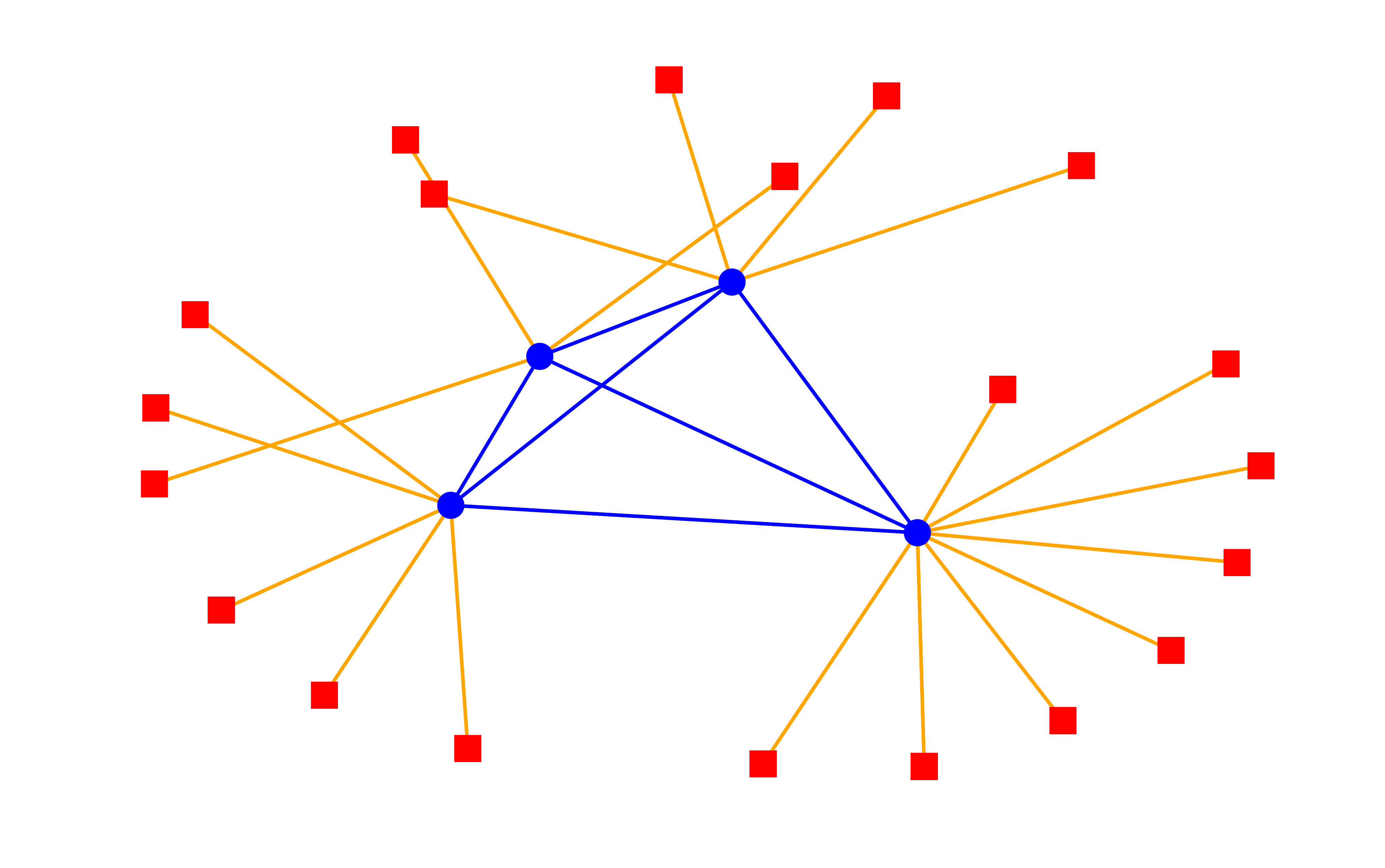}
}%subfigure
\caption{Evolution of the traffic load over time (a); mesh-like (b) and star-like (c) network topologies, with red dots corresponding to RSUs and blue ones representing core switches.
}
\end{figure}

In SUMO, vehicles are associated with a random source and destination
locations on the edge of the road topology, and move from the former
to the latter following the fastest (not necessarily the shortest)
route. The mobility simulated by SUMO accounts for such factors as
speed limits on different roads, the number and direction of lanes
therein, vehicles altering their course to overtake and/or avoid
incidents, and traffic lights. The resulting average and maximum
speeds are~12.9 and~70.1~km/h respectively, while the average
acceleration and deceleration values are~0.44 and~0.33~m/s$^2$. The
maximum deceleration, corresponding to vehicles violently braking to
avoid an obstacle, is 128.8~m/s$^2$. 
Both the speed and density of vehicles in the trace, depicted in \Fig{trace-speed} and \Fig{trace-density}, respectively, closely reproduce their real-world counterparts, with higher speeds along the main thoroughfares and higher densities around busy intersections.

The topology also includes 20 RSUs,
represented by red dots in \Fig{roads}.
RSUs are placed at the busiest road intersections, so as to cover a large set of vehicles. Specifically, we employ the following greedy procedure for RSU placement:
\begin{enumerate}
    \item we consider a set of {\em candidate locations};
    \item for each candidate location, we compute a {\em score}, corresponding to the number of vehicles passing through it;
    \item we place one RSU at the candidate location with the highest score;
    \item we subtract the newly-covered vehicles from the scores of all candidate locations;
    \item we repeat steps 3--4 until all RSUs are placed.
\end{enumerate}
While more complex deployment strategies
exist~\citep{malandrino2013optimal,liang2012optimal}, they are
typically tailored around one specific application, while we are
interested in modeling general-purpose vehicular networks supporting
several services.

RSU coverage and interference are computed according to the model presented in~\citep{trace}, which results in a maximum coverage radius of 255~meters. On average, successfully-transmitted beacons travel 123~meters between vehicles and RSUs.

At any given time, there are between~1,000 and~2,500 vehicles present in
the topology,
a value representative of the morning and afternoon peak
  times (i.e., 8:00-8:30 am and 5:00-5:30 pm, respectively.)
All vehicles send a beacon each
second~\citep{beacons-falko,avip}, which yields the traffic demand
depicted in \Fig{demand}. Notice that, while vehicles not covered by an
RSU still generate beacons, those beacons do not reach the collision
detectors, and thus are not accounted for in \Fig{demand}.

Our real-world trace contains no information on the network topology,
i.e., how the RSUs are connected with each other and with the core
network. Network topologies can have a substantial impact on
performance; intuitively, we can expect sparser topology to put a higher
stress on switches -- and the controller controlling them. We study two
such topologies, represented in \Fig{network-topo-mesh} and
\Fig{network-topo-star} respectively. In both topologies,
we create one switch for each RSU (red dots in the figure), and add four
core-level switches (blue dots).
In the mesh-like topology (\Fig{network-topo-mesh}), we
then connect:
\begin{itemize}
    \item the core switches in a mesh (blue links in \Fig{network-topo-mesh});
    \item each RSU switch to the two closest core switches (orange links in \Fig{network-topo-mesh});
    \item each RSU switch is also connected to the two closest RSU switches (black links in \Fig{network-topo-mesh}).
\end{itemize}
The star-like topology, shown in \Fig{network-topo-star}, is less connected. With respect the mesh-like topology:
\begin{itemize}
    \item RSU switches are only connected to the closest core switch;
    \item there are no links between RSU switches.
\end{itemize}

In our experiments, the number and location of collision detectors is
not determined {\em a priori}: collision detectors can be placed at any
RSU or core switches. We {\em refine} these decisions through the
greedy, iterative process described later in  \Sec{placement}.
Furthermore, we remark that the above 24 switches are controlled by a
single SDN controller.  

\subsection{Mininet network structure}
\label{sec:testbed-structure}

\begin{figure}[h]
\centering
\includegraphics[width=.5\textwidth]{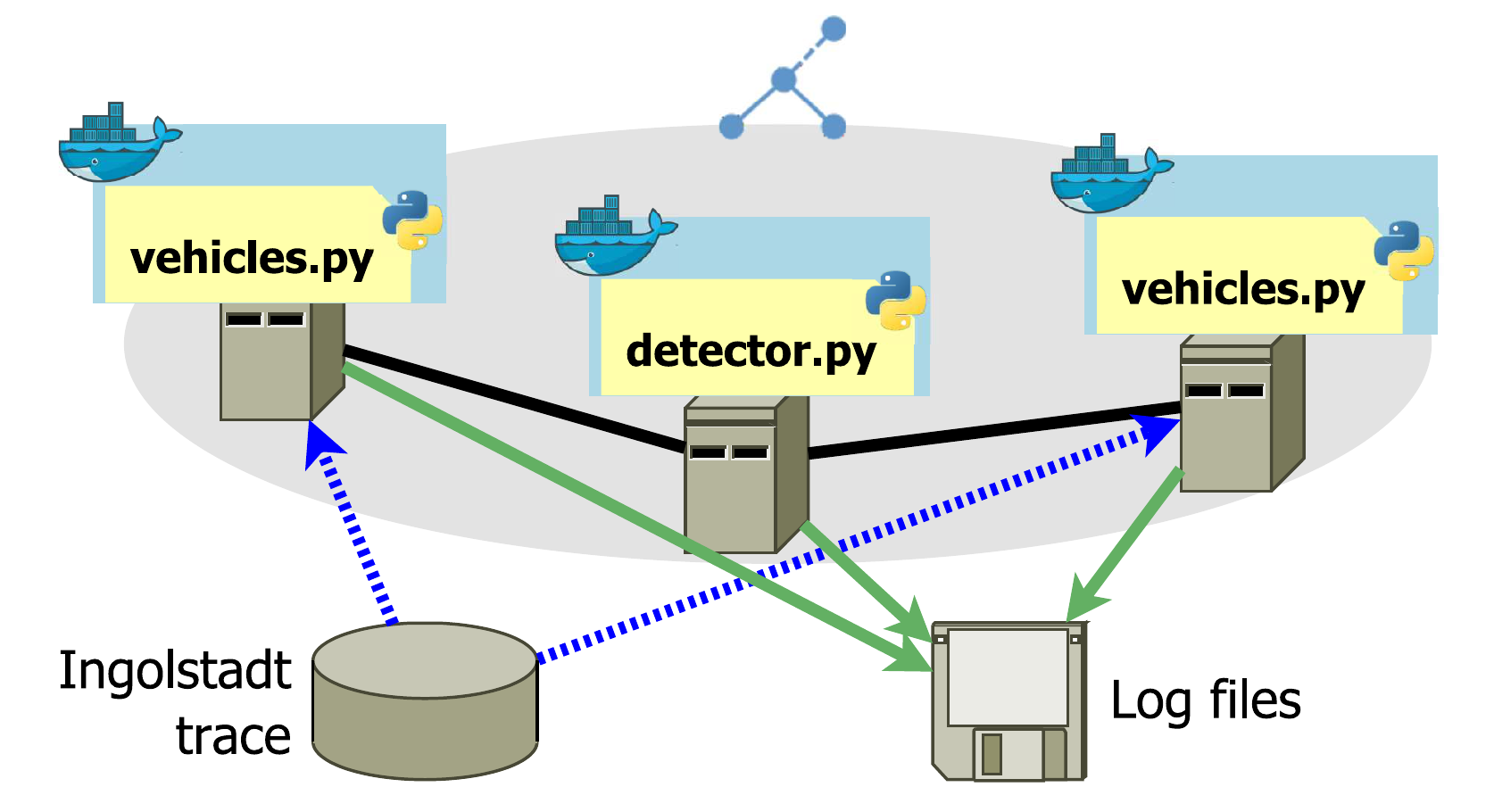}
\caption{
Our network architecture. A Mininet network (the gray area) contains 
several Mininet hosts (switches are not represented for simplicity). 
Within each host, we run a Docker container, and within the container 
one of two Python scripts: \protect\path{vehicles.py} emulates the 
vehicles passing by an RSU, while \protect\path{detector.py} is a collision detector.
Mobility information is read from the Ingolstadt trace described in \Sec{mobility}. 
Both collision detectors and vehicles store detailed log information, 
which is later used to obtain the performance metrics presented in \Sec{results}.
\label{fig:testbed}
} %caption
\end{figure}

The basic structure of our network is summarized in \Fig{testbed}. We
have a Mininet emulated network, including:
\begin{itemize}
    \item one OpenVSwitch controller, bundled with Mininet;
    \item one switch for each of the 20 RSUs and four extra {\em core} switches;
    \item one host per RSU;
    \item one host per collision detector.
\end{itemize}
Recall that switches are connected as described in
\Fig{network-topo-mesh}--\Fig{network-topo-star}. For Mininet-emulated links, we conservatively keep the
default bandwidth of 1~Gbit/s and the default latency of~0.12~ms.
Within each Mininet
host, we run a Docker container, and within each container we run a
Python program, which depends on the type of host (either a RSU or a
collision detector host).

As far as RSU hosts are concerned, we connect each host to the
corresponding RSU switch, and run the \path{vehicles.py}
script, which represents the vehicles under the RSU (see \Fig{testbed}), on the host. 
The \path{vehicles.py} script is in charge of:
\begin{itemize}
    \item reading the mobility information from the real-world trace
    from the city of Ingolstadt, described in \Sec{mobility};
    \item generating the beacons carrying the above mobility
    information, and transmitting them to the collision detector;
    \item receiving the replies from the collision detector, and logging
    the elapsed time.
\end{itemize}

The collision detector program, \path{detector.py} runs within the
collision detector hosts, and:
\begin{itemize}
    \item receives  beacons sent by the vehicles;
    \item detects  collisions, by running \Alg{collisions} described earlier;
    \item sends  collision reports as appropriate;
    \item logs the time it took to process each beacon.
\end{itemize}

Notice that, for each beacon, we log {\em two} times: the delay
perceived by the vehicle, i.e., the time elapsed between sending the
beacon and receiving the reply, and the time used by the collision
detector to actually process the beacon. The difference between these
times is the network delay, i.e., the time packets spend traveling from
the vehicle to the collision detector and vice versa within the emulated
network.

Each beacon/reply consists of a single UDP packet. Also, we stress that,
owing to the dynamic nature of vehicular scenarios, there are no
persistent connections between vehicles and collision detectors.

\noindent{\bf Controllers.}
SDN networks include a {\em controller}, a software program that determines 
the forwarding behavior of switches. In the simplest case:
\begin{itemize}
    \item switches have a set of {\em rules}, determining how packets
    shall be treated (forward on a certain port, flood, discard...);
    \item upon receiving a packet that does not match any of the
    existing rules, switches will forward it to the controller;
    \item based on the headers and/or payload of the packet, the
    controller will install one or more rules on the switch.
\end{itemize}
Being software programs, controllers can make switches behave in
virtually any way. One of the simplest behaviors controllers can
implement is the so-called {\em learning switch}: the controller
observes from which port of each switch packets coming from a certain
host are received, and ``learns'' that future packets directed to that
host shall be forwarded on the same port.

In our experiments, we compare two SDN controllers, both implementing
the learning switch behavior: Pox and Floodlight. Both are popular,
actively maintained open-source projects; however, they have slightly
different goals and scopes. Pox~\citep{pox} is written in Python and is
based on an older project called NOX; it aims at providing a simple,
object-oriented interface to OpenFlow, and is often used in research
projects. Floodlight~\citep{floodlight} is written in Java, and its
community tends to focus on providing high performance, configurability
(e.g., through a REST API) and manageability (through web-based GUIs).
Both controllers are vastly more capable than it is needed for our
scenario; however, we are interested to see whether they provide us with
different trade-offs between performance and complexity (hence, energy
consumption).

\subsection{Wireless simulations}\label{sec:wireless}

Since Mininet does not support the emulation of wireless networks, we cannot use it to study the delay incurred by beacons when going from vehicles to RSUs. Instead, we resort to {\em simulations}, based on the popular, open-source simulator ns-3~\citep{ns3}. ns-3 includes a detailed WAVE model, reproducing both its MAC layer and multi-channel coordination mechanism.

As specified by the IEEE~1609.4 standard, we set the control and
service channels (CCH and SCH, respectively) to take 50~ms
each. All beacons are transmitted on the CCH and all communication happens on the 5.9-GHz band, with a channel bandwidth of~10~MHz.
We perform our simulations as follows:
\begin{itemize}
    \item we take the position of RSUs and the mobility of vehicles from the Ingolstadt trace described in \Sec{mobility}, so as to guarantee the consistency between simulation and emulation;
    \item as in the emulated scenario, vehicles transmit a beacon every second;
    \item we measure the time it takes for beacons to reach the RSUs.
\end{itemize}
We then add the {\em beacon-specific} delays we obtain from the
simulation to our emulation results, thus being able to account for
the radio link contribution to the total latency.

\section{Collision detector placement}
\label{sec:placement}

As mentioned in \Sec{mobility}, a key feature of our scenario is that
{\em any} number of collision detectors can be attached at {\em any}
point of the network topologies described in
\Fig{network-topo-mesh}--\Fig{network-topo-star}. This reflects the
increased flexibility offered by the network function virtualization
(NFV), where any network node can run (virtually) any program. We
therefore have to establish (i) {\em how many} collision detectors we
need in our network in order to ensure that a sufficiently high fraction
of beacons  are served within the deadline set by the application, and
(ii) {\em where} in the network topology these detectors should be
placed.

Assuming we want at most one detector per node, this translates into
deciding, for each of the 24 network nodes depicted in
\Fig{network-topo-mesh}--\Fig{network-topo-star} (20 RSUs plus 4 core
switches), whether or not we place a detector therein. This produces a
total of $2^{24}\approx 16\cdot 10^6$~combinations. Recall that, because
we are {\em emulating} networks, as opposed to simulating them, testing
one combination with the one-hour trace we use also takes one hour.
Thus, testing all possible combinations is clearly impractical. A
popular and effective approach is coupling network simulation (or
emulation, in our case) with stochastic optimization algorithms, as done
in~\citep{hess2012optimal}. Intuitively, stochastic optimization
techniques~\citep{genetic} are based on evaluating the performance ({\em
fitness}) of a set of randomly-generated solutions, combining the most
promising ones into new solutions to evaluate, and repeating the process
until convergence is reached. They have been shown~\citep{annealing} to
find optimal or quasi-optimal solutions after testing a very limited
number of alternatives, i.e., performing a very limited number of
simulations (or emulations in our case).

Considering that optimization is not the focus of our study, we further
simplify the collision detector placement, and follow the {\em greedy
refinement} procedure below. Given the number~$n$ of detectors to
deploy, we:
\begin{enumerate}
    \item start by placing the detectors at randomly chosen nodes;
    \item emulate the configuration thus obtained, and consider, for
    each RSU, the {\em success fraction}, i.e., the fraction of beacons
    originated within the RSU coverage for which a reply from the
    collision detector is received within the deadline;
    \item for each switch (either RSU or core switch), compute the
    success fraction corresponding to the neighboring RSUs;
    \item move a detector from the switch with the highest success
    fraction (among those having a detector) to the switch with the
    lowest success fraction (among those not having a detector);
    \item if the configuration has been already tested, move a
    randomly-chosen detector to a randomly-chosen switch;
    \item go to step 2.
\end{enumerate}
Notice how the random changes in step~5 are equivalent to the mutation
step in genetic~\citep{genetic} and simulated annealing~\citep{annealing}
algorithms. Furthermore, a desirable aspect of our procedure is that
there are no meta-parameters that need tweaking: this simplifies our
study, and guarantees that none of the results we will observe is an
artifact of a specific parameter setting.

Although we cannot formally prove any property in this respect, we
consistently observed the greedy refinement procedure outlined above to
converge in twenty to thirty iterations, corresponding to an emulation
time of roughly one day. Additionally, the runs for different values
of~$n$ are independent and can be run in parallel: indeed, all the
results we show in \Sec{results}  can be obtained over a weekend.

\section{Numerical results}
\label{sec:results}

For our performance evaluation, we set the deadline by which replies
shall be received to~20~ms,
as suggested by the real-world motorway trial~\citep{digitala9}.
Then, in
\Sec{results-default}, we change the number~$n$ of detectors between~1 and~5 and, for
each value of~$n$, we study the overall
detection performance, e.g., the fraction of successfully processed
beacons,
along with the associated delay and energy consumption. 
In \Sec{results-scenarios} we investigate  how
changing the core network topology or the SDN controller influences the
system performance and energy consumption. 
Finally, \Sec{results-koeln}
presents some results obtained using a different road topology and
user mobility trace.

\begin{figure}[h]
\centering
\subfigure[\label{fig:performance}]{
    \includegraphics[width=.46\textwidth]{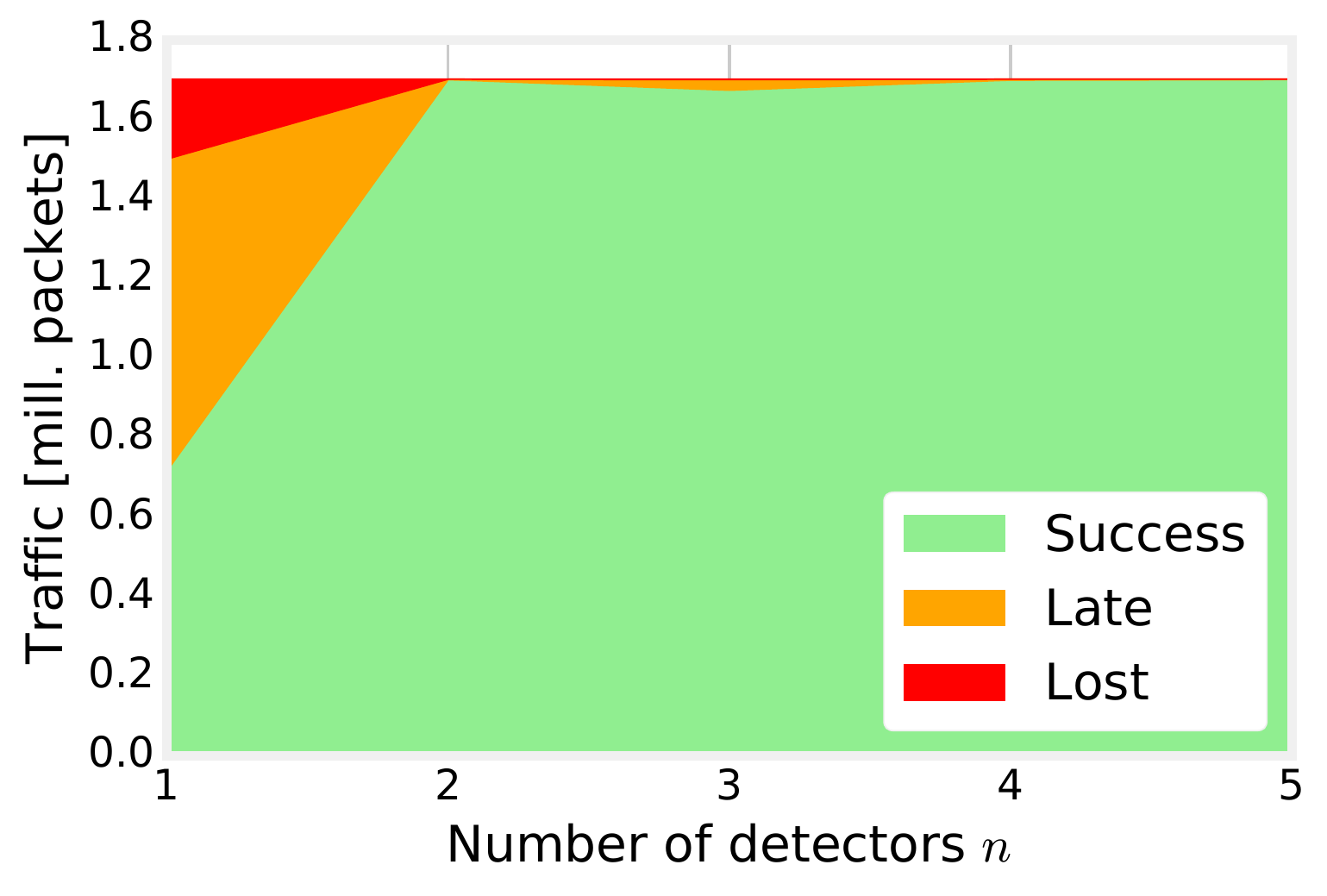}
}%subfigure
\subfigure[\label{fig:in-delay}]{
    \includegraphics[width=.46\textwidth]{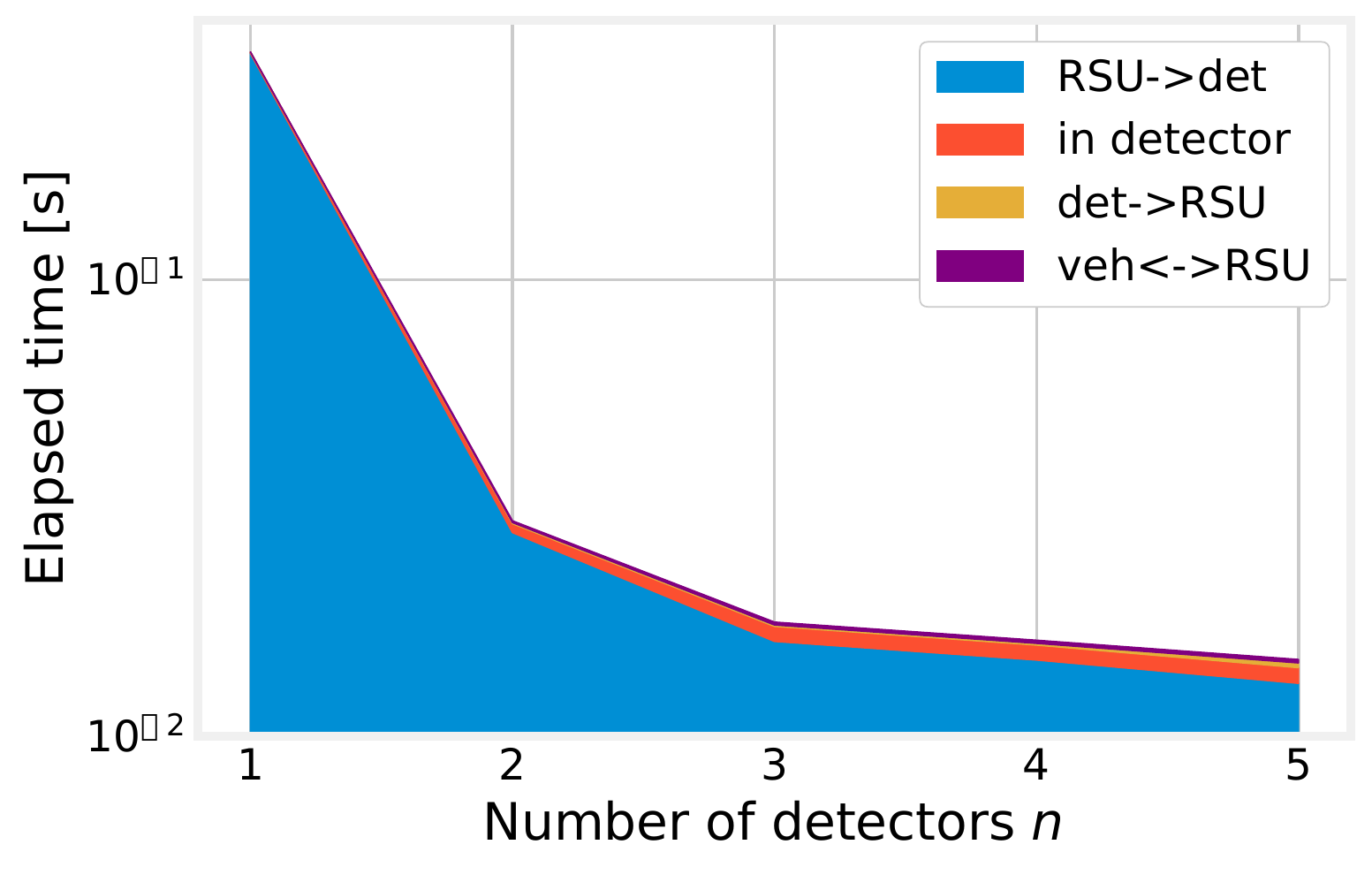}
}%subfigure
\\
\subfigure[\label{fig:delay-cdf}]{
    \includegraphics[width=.46\textwidth]{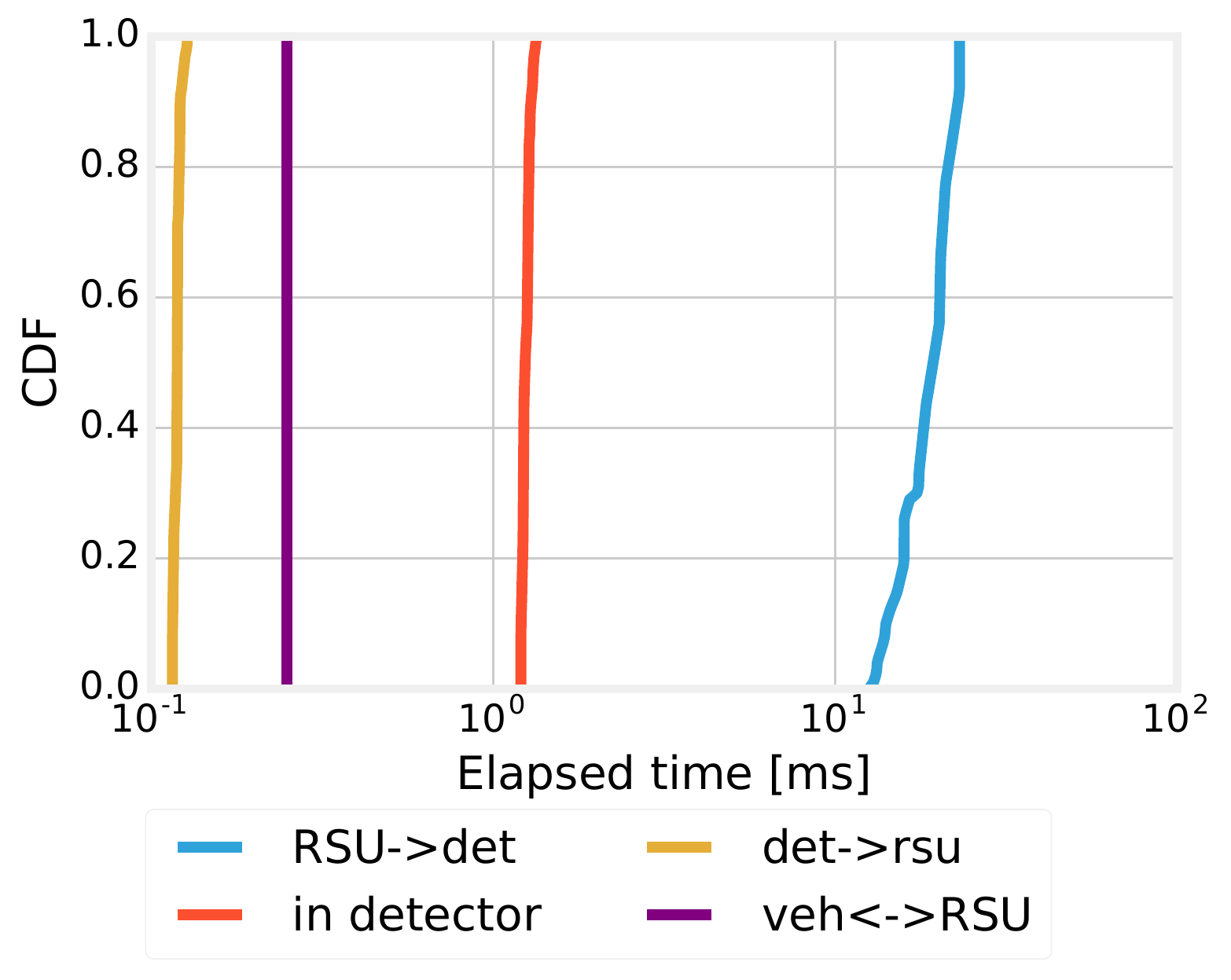}
}%subfigure
\caption{Default scenario. (a): Number of successfully processed, delayed and 
lost beacons as a function of the number~$n$ of detectors. (b): Breakdown 
of the delay in its components. (c): Distribution of the delay components when~$n=2$.
}
\end{figure}

\subsection{Default scenario}
\label{sec:results-default}

In the following, we consider a {\em default} scenario, where:
\begin{itemize}
\item the road topology and user mobility are modeled as described in \Sec{mobility};
    \item we use the star-like core network topology depicted in
    \Fig{network-topo-star};
    \item all switches are controlled by a Pox~\citep{pox} SDN controller.
\end{itemize}
The most basic aspect we are interested in is the effectiveness of our
collision detection system. Out of all the beacons sent by vehicles, we
need to know how many are (i) {\em successfully} processed, i.e.,
receive a response within the set deadline; (ii) {\em late}, i.e.,
receive a response but later than the deadline; (iii) {\em lost}, i.e.,
never receive a response. These three cases are represented by green,
yellow and red areas in \Fig{performance} respectively.

\Fig{performance} shows that, as long as there is more than one collision 
detector deployed in the network, virtually all beacons can be processed 
within the deadline. Only in the case~$n=1$ we can observe a small number 
of lost beacons, and a substantial fraction of beacons that are replied to 
too late. Bearing in mind
that we are taking into account a medium-sized European city under
congested traffic conditions, our results suggest that the task of
collision detection can indeed be successfully tackled through a
vehicular network based on SDN/NFV.

\Fig{in-delay} breaks the delay down into four components:
\begin{itemize}
    \item the time to reach the detector from the RSU (labeled as~\path{RSU->det});
    \item the processing time within the detector, e.g., to run \Alg{collisions} (labeled as~\path{in_detector});
    \item the time to reach RSU from the detector (labeled as~\path{det->RSU});
    \item the time beacons and alerts spend in the air (labeled as~\path{veh<->RSU}).
\end{itemize}
While we might expect these components to be roughly equivalent,
\Fig{in-delay} shows the opposite: the time to transfer the beacons from
the RSUs to the detector outweighs all other components; as confirmed by
the CDFs in \Fig{delay-cdf}, the difference is of almost two orders of
magnitude. Interestingly, \Fig{delay-cdf} also shows that the time
needed by data to travel in the opposite direction, i.e., from the
detector to the RSUs, is much shorter, even shorter than the processing
time at the detector.

\begin{figure}[h]
\centering
\subfigure[\label{fig:energy-tot}]{
    \includegraphics[width=.46\textwidth]{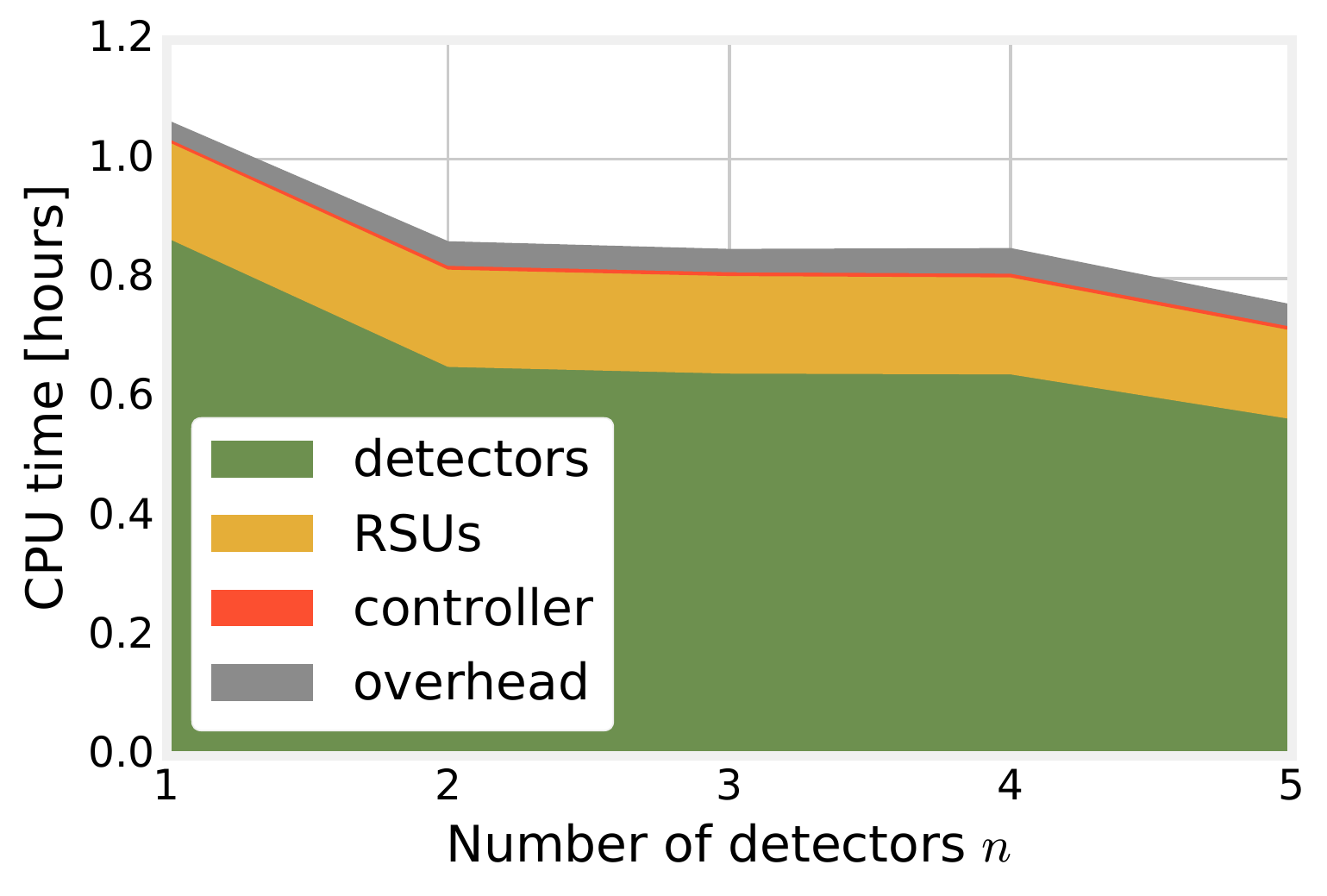}
}%subfigure
\subfigure[\label{fig:energy-det-traffic}]{
    \includegraphics[width=.46\textwidth]{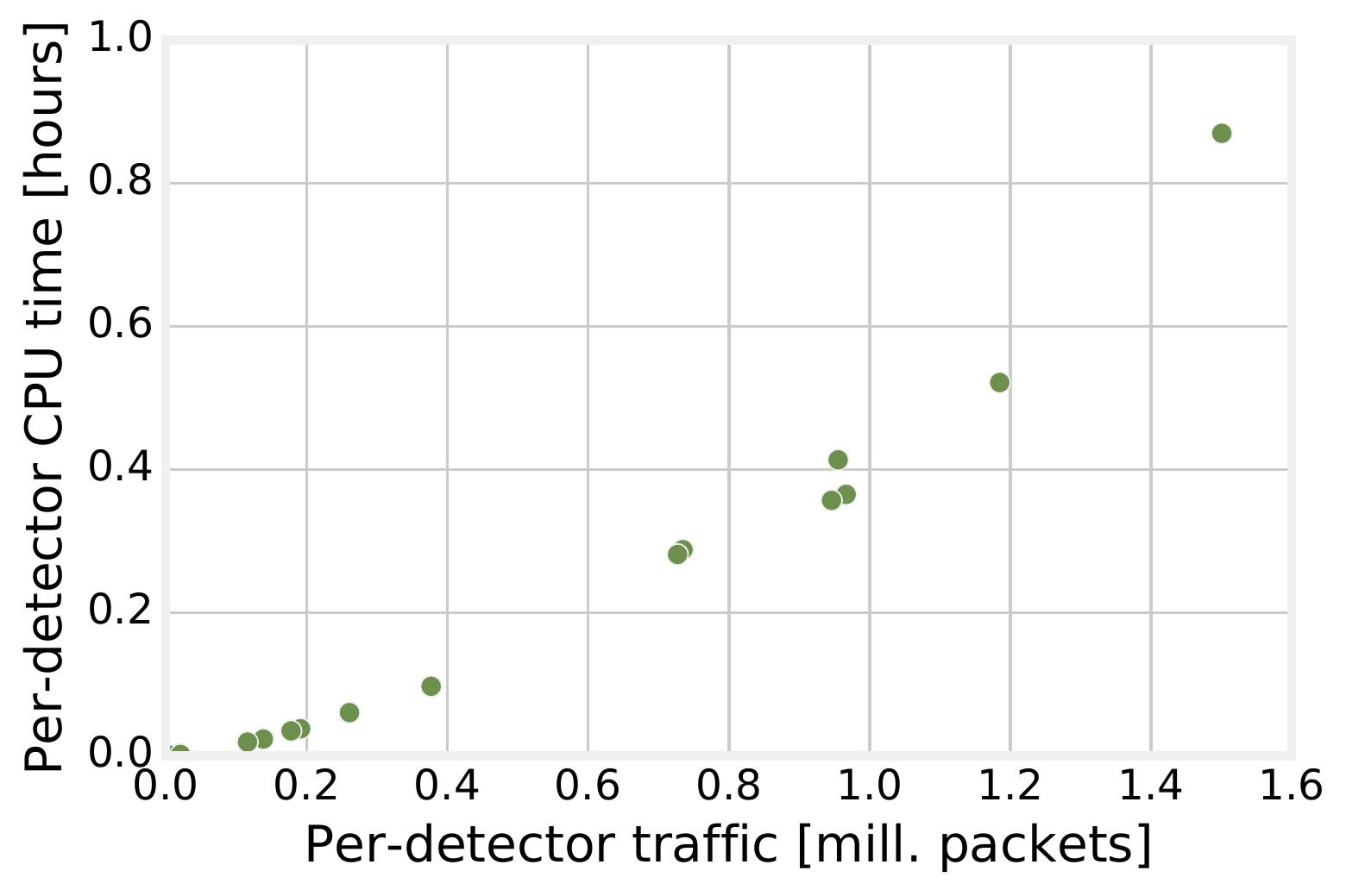}
}%subfigure
\\
\subfigure[\label{fig:energy-det-served}]{
    \includegraphics[width=.46\textwidth]{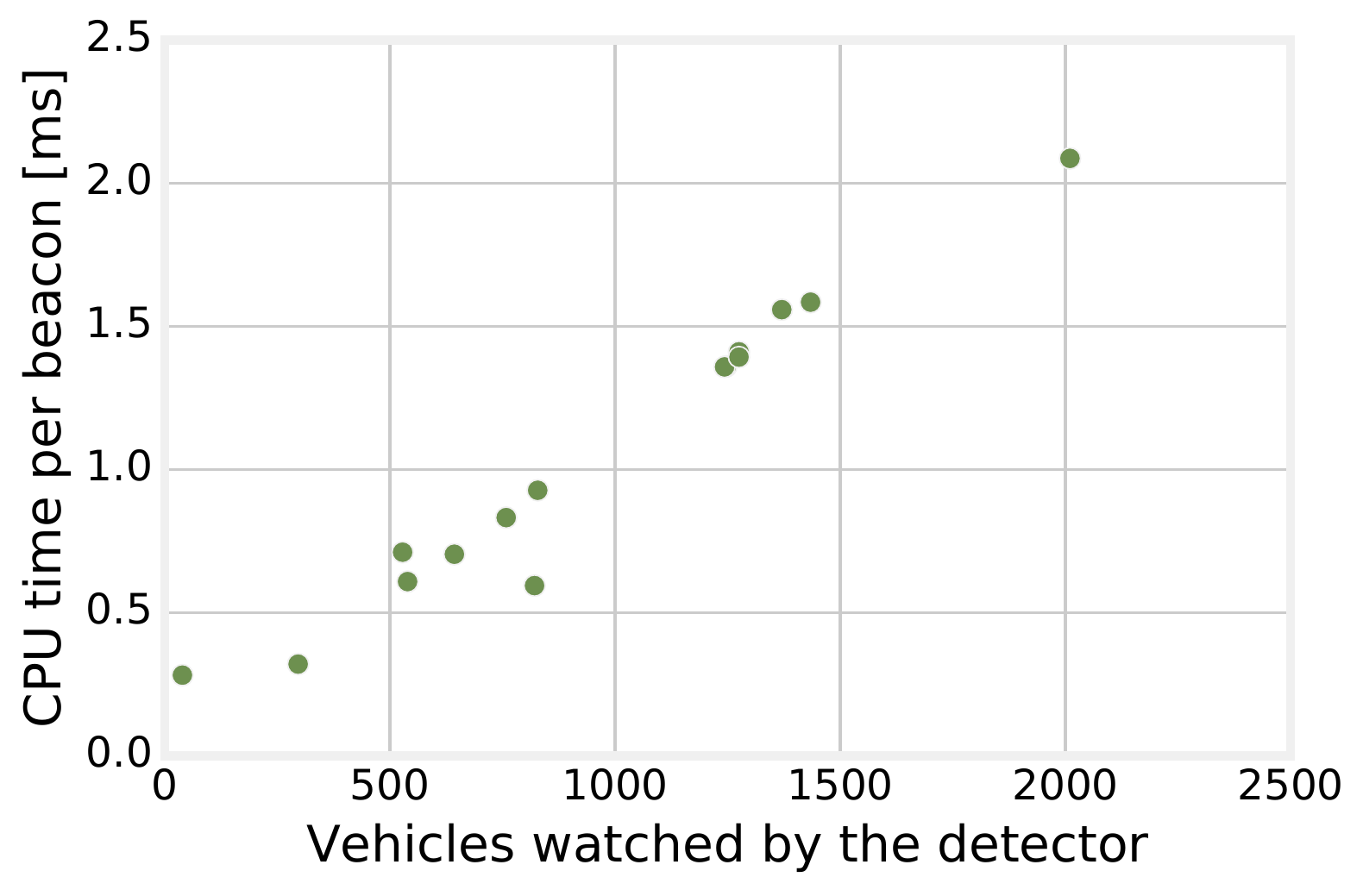}
}%subfigure
\caption{Default scenario. (a): How much CPU time is consumed by the \protect\path{detector.py} 
script simulating detectors (yellow), \protect\path{vehicle.py} scripts simulating vehicles (green), 
the \protect\path{pox} controller (red), Mininet and Docker overhead (gray), for each value of~$n$. (b): 
Link between the number of beacons processed by a detector and the CPU time it consumes. (c): 
Link between the number of vehicles watched by each detector and its per-beacon CPU consumption.
\label{fig:energy}
}
\end{figure}

This is due to the fact that, while most packets are directly processed
at the switches, some -- those that do not match the forwarding rules
currently stored at the switch -- are forwarded to the SDN controller,
which substantially increases the amount of time needed to forward the
packets. Indeed, the OpenVSwitch virtual switches first cache the forwarding
instructions of the SDN controller for some time (which explains why the
replies going from the detectors to the RSUs are much less likely to be
forwarded to the controller again) and then purge them after a timeout,
in order to avoid keeping stale routes.

This unexpected effect serves us as a reminder that SDN does not
represent a drop-in replacement for traditional networks, and special
attention ought to be devoted to the interaction between nodes of the
data plane and controllers. At the same time, it further highlights how
network emulation is an excellent tool to study SDN networks.

In \Fig{energy}, we move to energy consumption. Specifically, we use the
CPU time logged by the different components of our system as a proxy for
the actual energy they consume; this is in line with such recent works
as~\citep{krishnan}, that identify an almost-linear relationship between
CPU utilization and energy consumption.

\Fig{energy-tot} shows the CPU time logged by detectors (i.e., the
\path{detector.py} instances), RSUs (i.e., \path{vehicle.py} instances)
and controllers, as a function of the number~$n$ of detectors. It also
represents the overhead due to Mininet, Docker, and the virtual machine
Mininet runs on (gray area in the plot). Recall that our tests last
one hour, and the total consumed CPU time can exceed that because
different components, e.g., two collision detectors, can use different
CPUs at the same time.

A first thing we can observe is that collision detectors consume most of
the CPU time; indeed, when~$n=1$, the detector is active for more than
50~minutes. \path{rsu.py} scripts also consume a fair amount of CPU, due
to their manifold role of sending the beacons, receiving the replies,
and logging the elapsed times. The CPU time consumed by the detector, on
the other hand, is almost negligible, amounting to barely 30 seconds.
This confirms that SDN controllers {\em per se} do not substantially
increase the energy consumption of the networks they belong to, and SDN
itself is a suitable technology to use in energy-constrained network
scenarios.

Another interesting aspect we can learn from \Fig{energy-tot} is that
the total CPU time consumed {\em decreases} as~$n$ grows, even as the
system performance (\Fig{performance}) increases. To understand the
reason for this, we show in \Fig{energy-det-traffic} the CPU time
consumed by each detector as a function of the number of beacons it has
to process throughout the whole simulation. There are a total of fifteen
points in \Fig{energy-det-traffic}: one for the single detector deployed
when~$n=1$ (the topmost one, corresponding to the CPU time consumption
we see in the leftmost part of \Fig{energy-tot}), two for the two
detectors deployed when~$n=2$, and so on. We can clearly see that, the
more beacons a detector has to process, the more CPU time it will
consume.

\Fig{energy-det-traffic} is not especially surprising: detectors
basically run \Alg{collisions} every time they receive a beacon, so it
stands to reason that doing that more often translates into a higher CPU
consumption. More interestingly, \Fig{energy-det-served} correlates the
{\em per beacon} CPU consumption with the number of vehicles each
detector has within its coverage area. We can observe an almost linear
correlation between the two. It tells us that having more vehicles to
deal with not only means that collision detectors need to process more
beacons, but also that each beacon takes longer to process. The reason
lies in the structure of \Alg{collisions} itself: in \Line{forb}, we
loop over all (recent) beacons received by other vehicles, and the
number thereof directly depends upon the number of vehicles the detector
has to watch.

\noindent{\bf Summary.}
In our default scenario (star-like topology as depicted in
\Fig{network-topo-star}, Pox controller), any value of~$n$ greater than
one guarantees that virtually all beacons are processed successfully
(\Fig{performance}). Having to send some packets to the SDN controller
is the main source of delay (\Fig{in-delay}, \Fig{delay-cdf}), and
collision detectors consume most of the CPU time (\Fig{energy-tot}), and
thus most of the energy. Such a consumption increases with the total
traffic each detector has to process (\Fig{energy-det-traffic}), as well
as the number of vehicles it has to watch (\Fig{energy-det-served}).
This suggests that improved, more efficient collision detection
algorithms are a worthwhile direction to follow in order to reduce the
energy consumption of vehicular safety networks.

\subsection{Alternative backhaul topology and detector}
\label{sec:results-scenarios}

\begin{comment}
\begin{figure*}
\centering
\subfigure[\label{fig:where-number}]{
    \includegraphics[width=.3\textwidth]{img/itcd1_where_detectors}
}%subfigure
\subfigure[\label{fig:where-ndist-star}]{
    \includegraphics[width=.3\textwidth]{img/itcd2_dist_treelike}
}%subfigure
\subfigure[\label{fig:where-ndist-mesh}]{
    \includegraphics[width=.3\textwidth]{img/itcd3_dist_meshlike}
}%subfigure
\caption{Mesh-like topology (``mesh'') and Floodlight controller (``floodlight'') scenarios. (a): Number of successfully processed, delayed and lost beacons as a function of the number~$n$ of detectors. (b): Breakdown of the delay in its components. (c): Distribution of the delay components when~$n=2$.
\label{fig:where}
}
\end{figure*}
\end{comment}

\begin{figure}[h]
\centering
\subfigure[\label{fig:performance-var}]{
    \includegraphics[width=.46\textwidth]{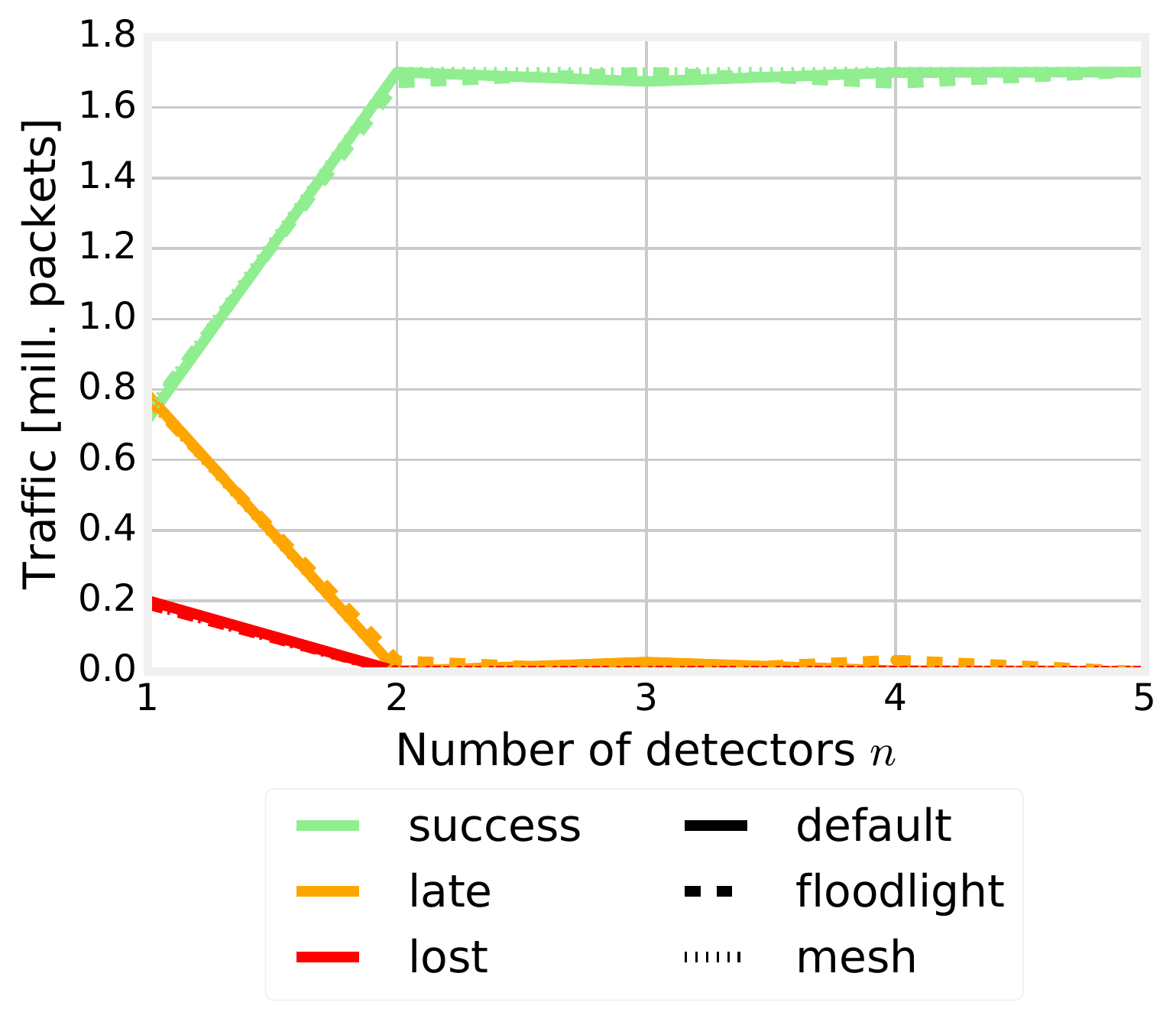}
}%subfigure
\subfigure[\label{fig:meandelay-var}]{
    \includegraphics[width=.46\textwidth]{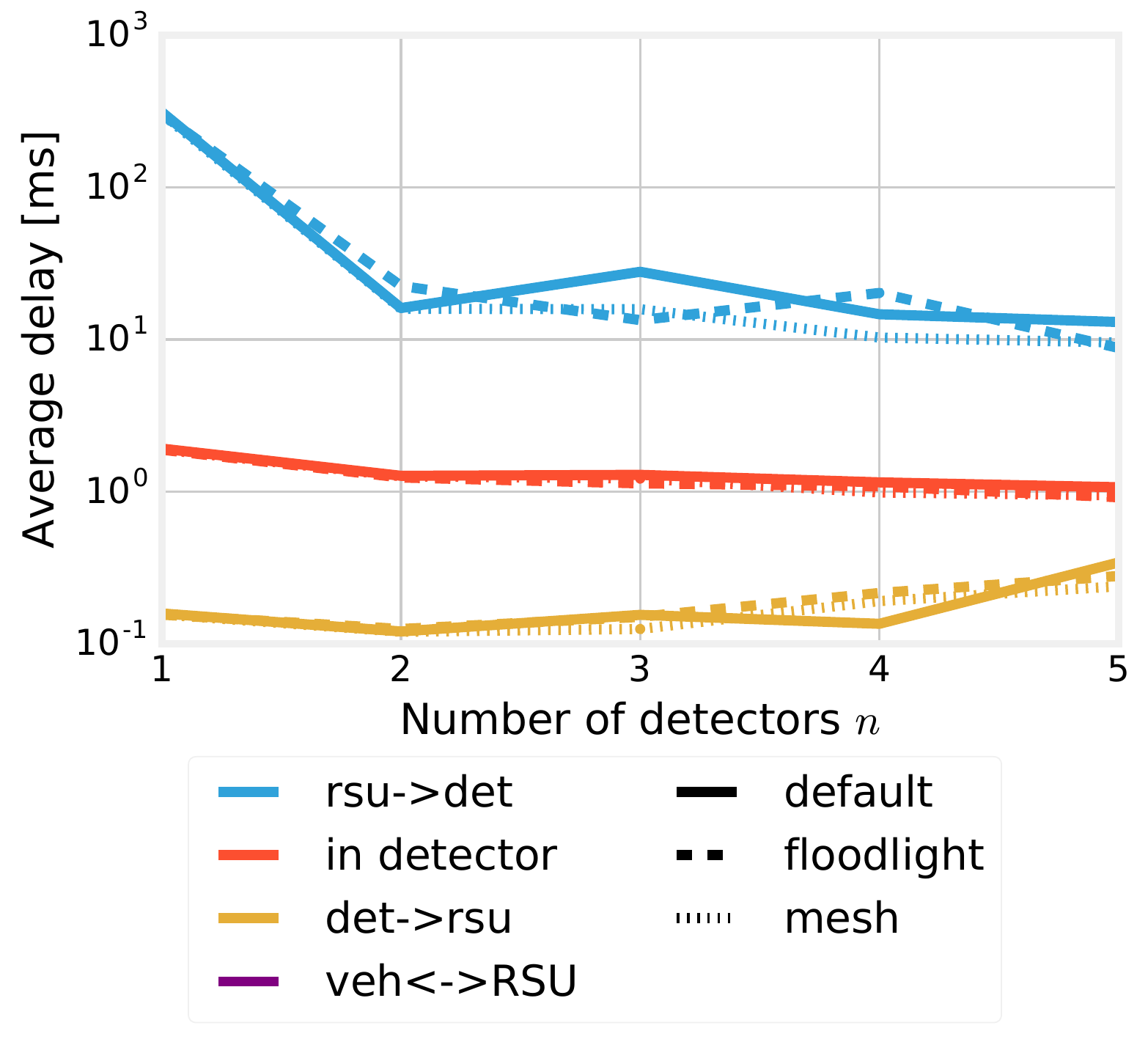}
}%subfigure
\\
\subfigure[\label{fig:delaycdf-var}]{
    \includegraphics[width=.46\textwidth]{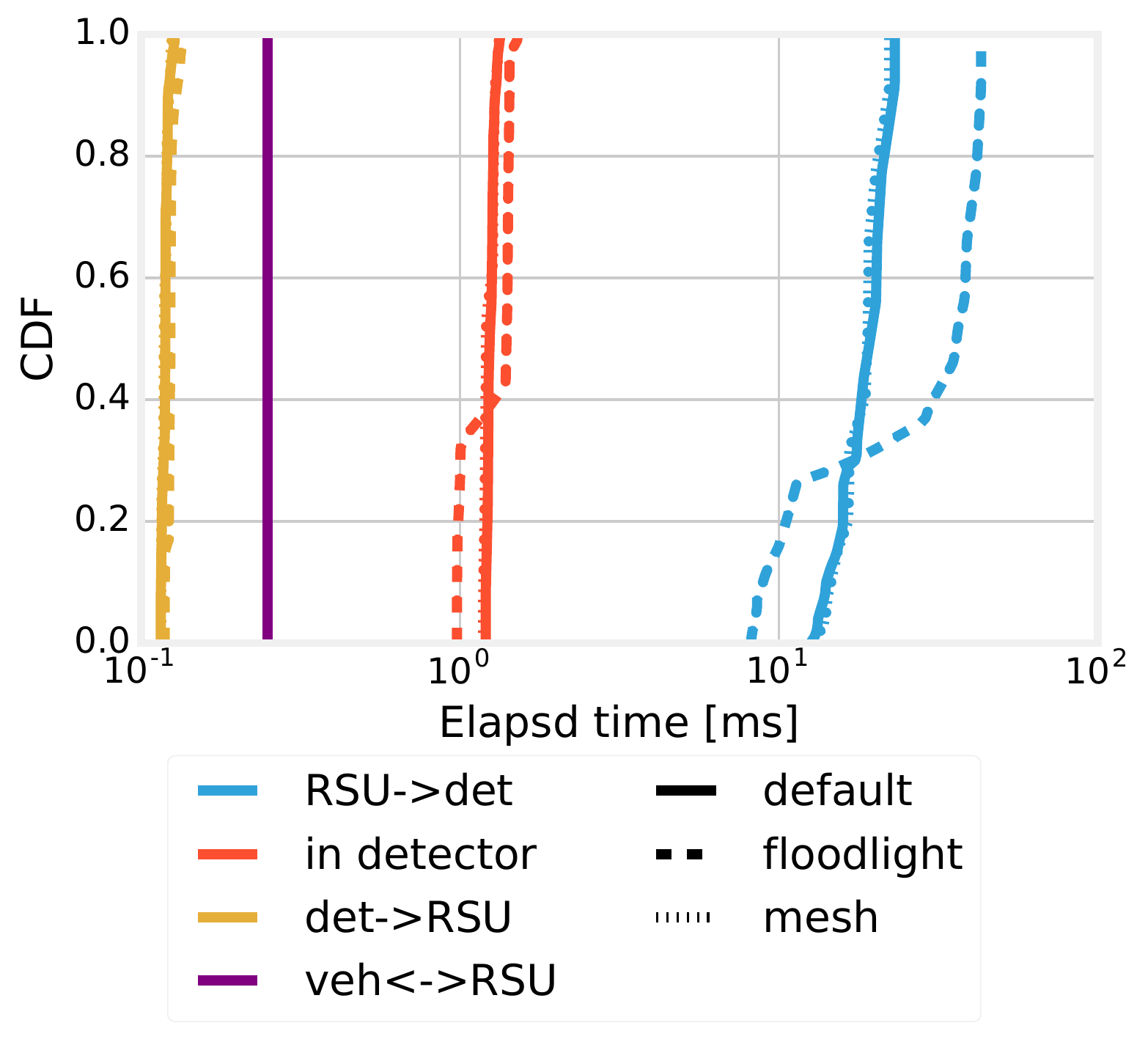}
}%subfigure
\caption{Mesh-like topology (``mesh'') and Floodlight controller (``floodlight'') scenarios. 
(a): Number of successfully processed, delayed and lost beacons as a function of the number~$n$ 
of detectors. (b): Breakdown of the delay in its components. (c): Distribution of the delay 
components when~$n=2$.
\label{fig:varscenario}
}
\end{figure}

In the following, we maintain the same road topology and mobility
trace as considered before, and address two alternative backhaul scenarios:
\begin{itemize}
    \item one labeled ``mesh'', where we replace the star-like network
    topology depicted in \Fig{network-topo-star} with the mesh-like one
    depicted in \Fig{network-topo-mesh};
    \item one labeled ``floodlight'', where we replace the Pox
    controller with the Floodlight~\citep{floodlight} one.
\end{itemize}
Notice that we are interested in studying the effect of these two
changes individually; therefore, in the ``mesh'' scenario we use the
same Pox controller as in the default one, and in the ``floodlight''
scenario we use the same star-like topology as in the default one.

\begin{figure}[h]
\centering
\includegraphics[width=.46\textwidth]{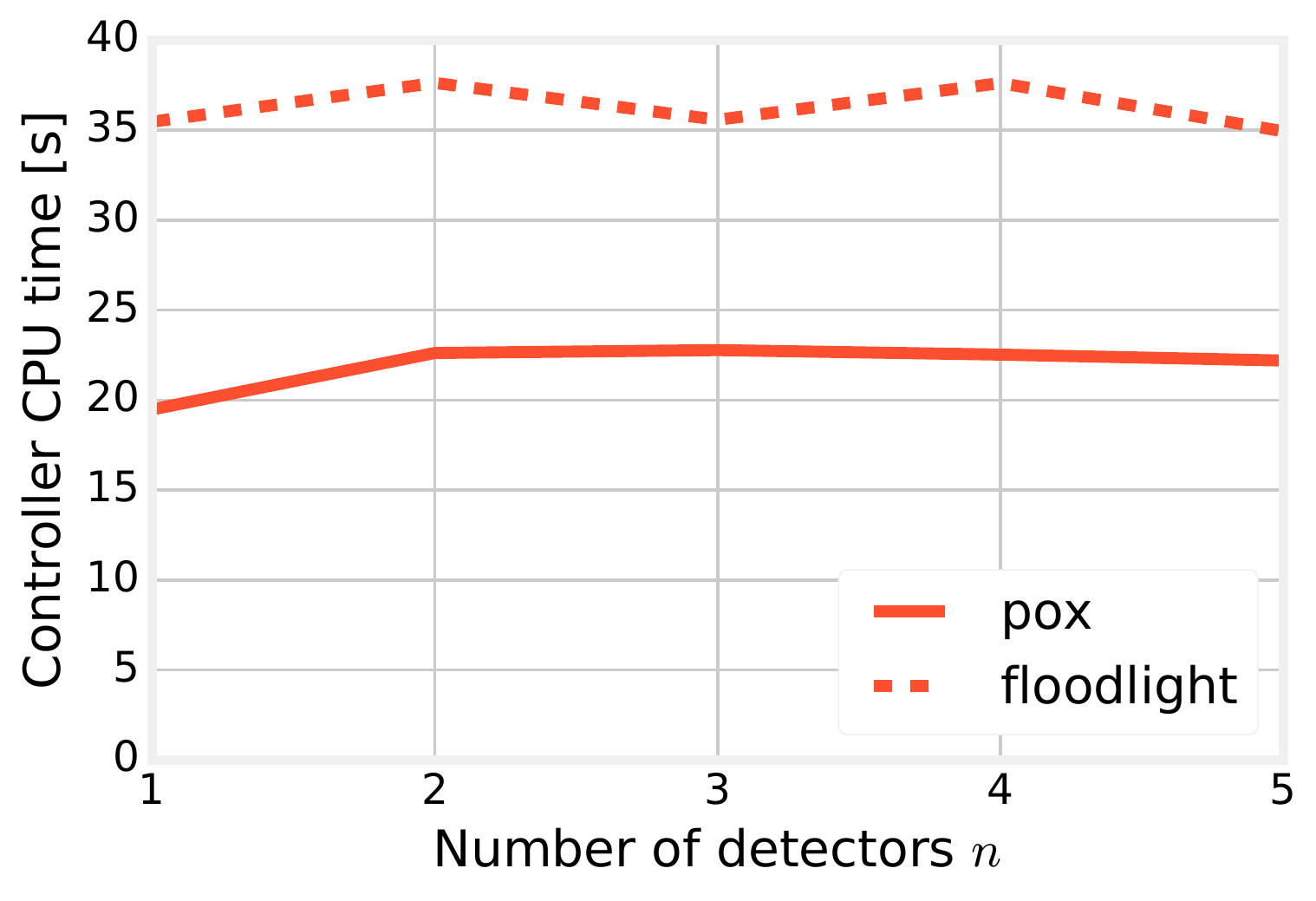}
\caption{
CPU time used by Pox (solid lines) and Floodlight (dashed lines) controllers, as~$n$ changes.
    \label{fig:controller-cpu}
}%caption
\end{figure}

\Fig{performance-var} shows that the performance is virtually the same
in all scenarios. Intuitively, this tells us that the collision
detection system we devised is robust to changes in the network topology
and type of SDN controller. There are, however, some slight but
significant differences in the delay: specifically, we can see from
\Fig{meandelay-var} that using the mesh-like network topology
corresponds to shorter delays, which again makes sense as in that case
individual switches tend to be less loaded.

More interestingly, in \Fig{delaycdf-var} we see that the Floodlight
controller is associated with a stronger variability in the delay,
especially for the packets sent from detectors to RSUs: some are
processed very quickly, while others take substantially longer than with
the Pox controller. Furthermore, this also affects the time spent by
packets {\em in} the controller (red lines in \Fig{delaycdf-var}), whose
variability increases as well. Indeed, as we observed earlier, the time
it takes the detector to process a beacon depends on how many beacons
the detector has received in the recent past, and that can change
substantially if controller-induced delays are not constant.

\Fig{controller-cpu} shows another difference between Pox and
Floodlight controllers: the latter consumes substantially more CPU time
than the former. Such a difference is due to the different language they
use (Java programs tend to be heavier than their Python counterparts),
and, to a greater extent, to Floodlight focusing on feature-richness over
simplicity.

\noindent{\bf Summary.}
Using a different controller or a different network topology does not
substantially change the system performance (\Fig{performance-var}).
However, a more connected topology translates into slightly shorter delays
(\Fig{meandelay-var}). Using the Floodlight controller {\em in lieu} of
Pox yields a higher variance in packet processing delay
(\Fig{delaycdf-var}), as well as higher CPU time consumption
(\Fig{controller-cpu}).

\subsection{Alternative road topology and mobility trace}
\label{sec:results-koeln}

\begin{figure}[h]
\centering
\subfigure[\label{fig:koeln-topo}]{
    \includegraphics[width=.46\textwidth]{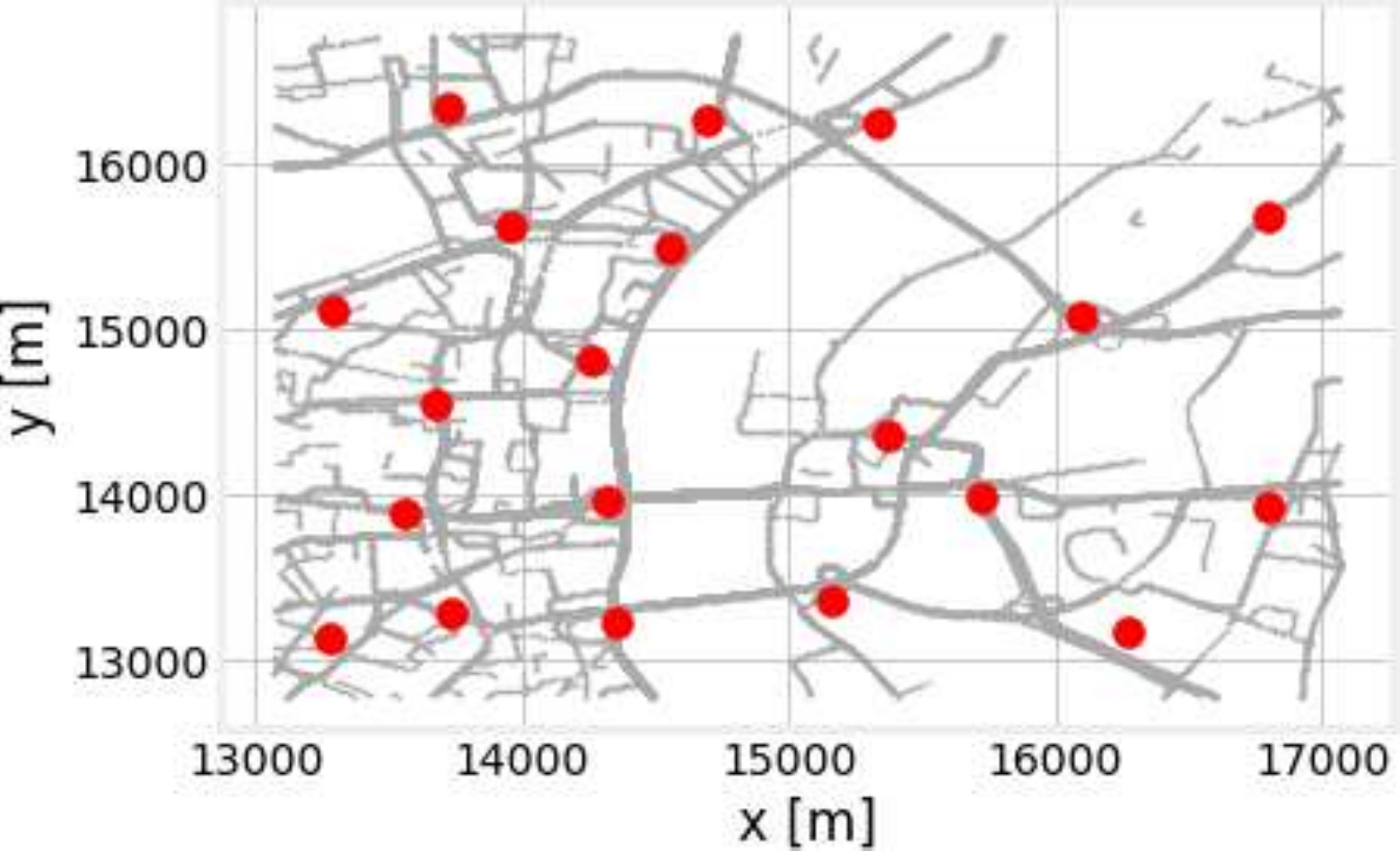}
}%subfigure
\subfigure[\label{fig:koeln-performance}]{
    \includegraphics[width=.46\textwidth]{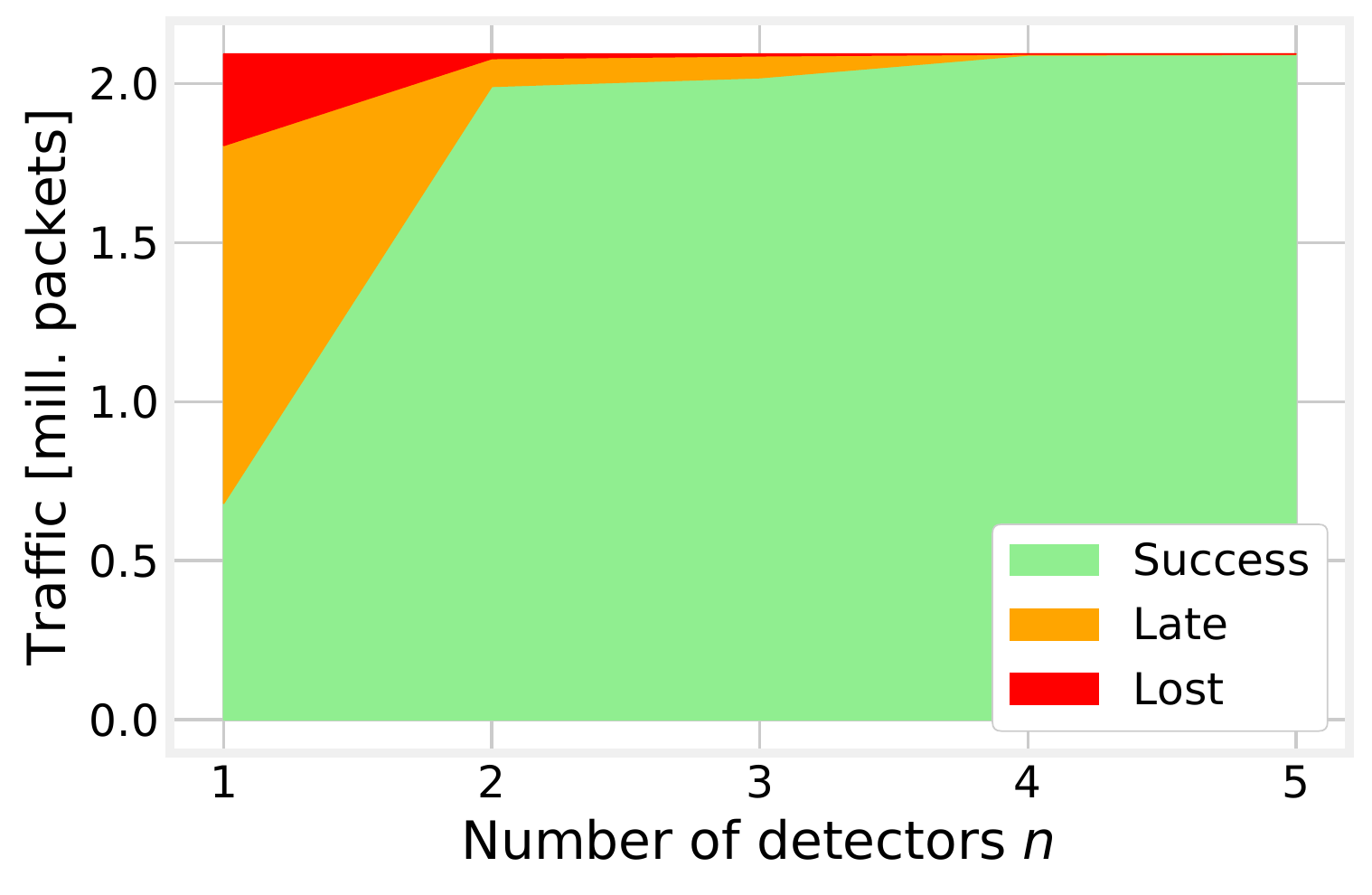}
}%subfigure
\\
\subfigure[\label{fig:koeln-delay}]{
    \includegraphics[width=.46\textwidth]{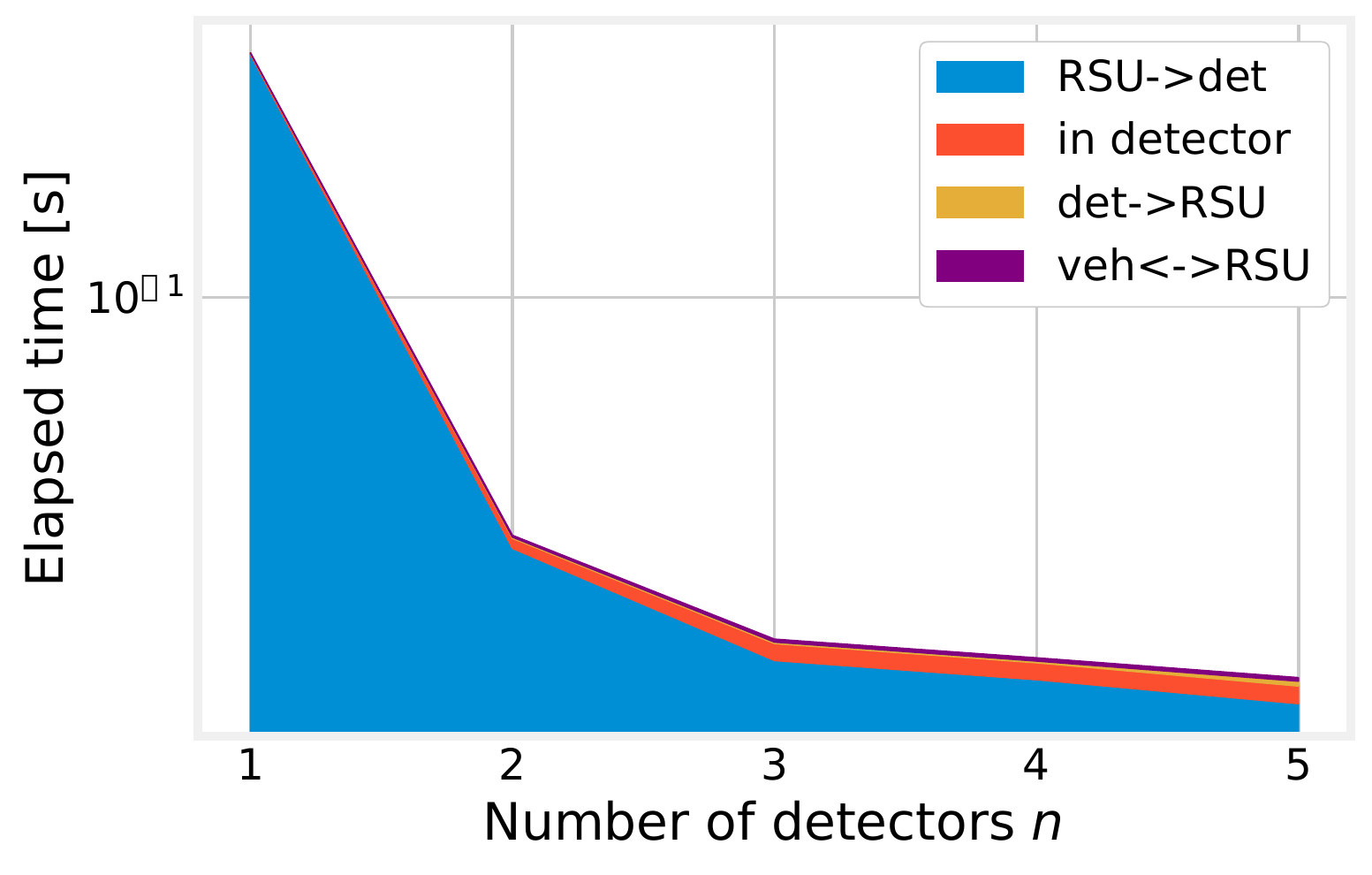}
}%subfigure
\caption{Cologne scenario. (a): road topology and RSUs location; (b): number of successfully processed, delayed and lost beacons as a function of the number~$n$ of detectors; (c): breakdown of the delay in its components.
\label{fig:varscenario}
}
\end{figure}

We now consider a different trace, coming from German city of Cologne~\citep{CologneTrace}. Similar to the Ingolstadt trace detailed in \Sec{mobility}, it combines real-world topology with realistic vehicle mobility obtained through SUMO. The area covered by the trace is 2~$\times$~2 km$^2$, and there are on average 2,410~vehicles, traveling at an average speed of 41.98~km/h. We place 20 RSUs on the topology, following the same greedy procedure as in the Ingolstadt case. \Fig{koeln-topo} shows the road topology (in gray) and the location of RSUs (red dots).

\Fig{koeln-performance} summarizes the number of beacons that are processed successfully, delayed, or lost. We can observe that, in spite of the higher number of beacons (notice the y-scale in the plot), two detectors are sufficient to provide the collision detection service with a small number of delayed or lost beacons.

We can further confirm this by comparing \Fig{koeln-delay} to
\Fig{in-delay}. 
Both the overall delay and its components are very similar between the
Ingolstadt and Cologne scenarios:   
the main difference lies in slightly longer delays in the collision
detectors for the Cologne scenario
(red area in \Fig{koeln-delay}), due to the higher number of vehicles.

\section{Related work}
\label{sec:relwork}

\subsection{Collision detection}

Collision detection for vehicular networks is a widely studied topic.
Earlier works such as~\citep{emergency} focus on system architecture,
e.g., the role of RSUs,
while later ones address specific aspects such as countering shadowing
effects~\citep{beacons-falko} or evaluating competing
systems~\citep{estimating}. In a recent twist, \citep{scarcrash}~advocates
using smartphone data along with the beacons that vehicles periodically
transmit.

Another significant aspect of collision detection systems is security.
Indeed, beacons can be used by malicious attackers to reconstruct the
vehicle position and/or trajectory~\citep{anonymity,pseudo}. Anonymous
beacons improve the situation~\citep{pseudo}; however, they can be abused
by vehicles providing false information~\citep{avip} to hide their
position to the authorities.

Compared to these works, the collision detection solution we present in
\Sec{collisions} is remarkably simple. This is due to the fact that our
focus is not on optimizing collision detection, but rather on assessing the
ability of SDN/NFV-based networks to meet the strict latency constraints
imposed by vehicular collision detection, and the resulting energy
consumption.

\subsection{SDN and NFV}

Our work is also related to the wide area of software-defined networking
and network function virtualization. In particular, the authors of the
early work~\citep{softcell} envision a software-based implementation of
next-generation cellular networks, where all types of network nodes,
e.g., firewalls and gateways, are implemented through {\em middleboxes},
virtual machines running on general-purpose hardware. The concept of
middleboxes is further generalized into virtual network functions
(NFV)~\citep{etsi-wp}, capable of performing any task, including those
usually carried out by ad hoc servers, e.g., video transcoding.

Enabled by SDN and NFV, mobile-edge computing (MEC) has been recently
introduced~\citep{fog} as a way to move ``the cloud'', i.e., the servers
processing mobile traffic, closer to users, thus reducing the latency
and load of networks. Recent works have studied the radio techniques
needed to enable MEC~\citep{mec-radio}, its relationship to the
Internet-of-things~\citep{mec-iot} and context-aware, next-generation
networks~\citep{mec-5g}.

Placing the VNFs and the servers hosting them within the cellular
network is one of the most important MEC-related research question, the
most popular approach being exact~\citep{placement-noi} and
approximate~\citep{placement-infocom,placement-cinesi} optimization. When
faced with the task of placing our collision detectors, we take the more
straightforward approach of refining their positioning, as detailed in
\Sec{placement}; indeed, for us the impact of different placement
solutions on the resulting delay and energy consumption is more
important than finding the utmost optimal solution.

\section{Conclusion and future work}
\label{sec:conclusion}

Collision detection is a prominent safety application of vehicular
networks, having very strict delay requirements. In order to verify the
compatibility of these requirements with SDN and NFV, we designed,
implemented and emulated one such collision detection system using
Mininet and Docker.

Using a real-world road topology and mobility trace, we found that a
limited number of collision detectors can process the vast
majority of beacons with acceptable delay. More importantly, we found
that most of that delay comes from packets being sent to the SDN
controller; this further highlights the importance of thoroughly testing
SDN-based solutions before deploying them.

\section*{Acknowledgement}
This work was partially supported by the European Commission through the 
H2020 5G-TRANSFORMER project (Project ID 761536).

\bibliographystyle{elsarticle-num}
\bibliography{refs}

\end{document}